\documentclass[a4paper,11pt]{article}     

\include{python_preamble}
\usepackage{jheppub}
\usepackage     [utf8]                  {inputenc}
\usepackage     [T1]                    {fontenc}
\usepackage                             {color}
\usepackage                             {amsmath}
\usepackage                             {amssymb}
\usepackage{bbold}
\usepackage                             {graphicx}
\usepackage     [english]               {babel}
\usepackage                             {hyperref}   
\usepackage{amssymb}
\usepackage{mathtools}
\usepackage{color}
\usepackage{multirow}
\usepackage{listings}
\usepackage{natbib}
\usepackage{tabularx}
\usepackage{tikz-feynman}
\usepackage{tikz}
\usetikzlibrary{shapes,arrows}
\usepackage{caption}
\usepackage{subcaption}
\usepackage{graphicx}
\usepackage{physics}
\usepackage{slashed}
\usepackage{cleveref}
\usepackage{longtable}
\usepackage{hhline}
\usepackage{ytableau}
\usepackage[normalem]{ulem}

\usepackage{soul}
\soulregister\cite7
\soulregister\ref7
\soulregister\eqref7
\soulregister\znubb7

\newcommand{\ovbb}{\text{$0\nu\beta\beta$}}
\newcommand{\ovbbp}{\text{$0\nu\beta\beta\phi$}}
\newcommand{\tvbb}{\text{$2\nu\beta\beta$}}
\newcommand{\psfm}{\text{$\Omega^{\phi}$}}

\newcommand{\thm}{\text{$T_{1/2}^{\phi}$}}

\newcommand{\Xe}{\text{${}^{136}$Xe}}
\newcommand{\Ge}{\text{${}^{76}$Ge}}
\newcommand{\param}{\text{$\boldsymbol{\theta}$}}
\newcommand{\psg}{\text{$\bar{g}$}}
\newcommand{\mix}{\text{$\mathcal{U}$}}
\newcommand{\vect}[1]{\mathbf{#1}}
\newcommand{\amp}{\text{$\mathcal{A}_\nu$}}
\newcommand{\stc}{\text{$g_{\phi N N}$}}
\newcommand{\stcg}{\text{$\bar g_{\phi N N}$}}

\newcommand{\xcheckmark}{\checkmark\kern-1.1ex\raisebox{.7ex}{\rotatebox[origin=c]{125}{--}}}

\title{Scalarful double beta decay}

\author[a,b]{Jordy de Vries,}
\author[b,c]{Luk\'{a}\v{s} Gr\'{a}f,}
\author[a,b]{Vaisakh Plakkot,}
\author[c]{and Dominik Star\'{y}\,}
\emailAdd{j.devries4@uva.nl}
\emailAdd{lukas.graf@matfyz.cuni.cz}
\emailAdd{v.plakkot@uva.nl}
\emailAdd{dominik.stary@matfyz.cuni.cz}

\affiliation[a]{Institute of Physics and Delta Institute for Theoretical Physics, University of Amsterdam,\\Science Park 904, 1098 XH Amsterdam, The Netherlands}
\affiliation[b]{Theory Group, Nikhef,\\ Science Park 105, 1098 XG, Amsterdam, The Netherlands}
\affiliation[c]{
Institute of Particle and Nuclear Physics, Faculty of Mathematics and Physics, Charles University in Prague,\\ V Holešovičkách 2, 180 00 Praha 8, Czech Republic}

\abstract{
    We revisit scalar emissions in double beta decays of nuclei, often discussed in the context of Majoron models, in light of the latest developments on the study of neutrinoless double beta decay amplitudes from an effective field theory approach. The sensitivity of double beta decay experiments to this process is assessed through an analysis of spectral shapes, and the study is extended to massive scalars, scalars coupling to sterile neutrinos, and exotic right-handed effective couplings.
}

\begin{document}

\maketitle

\section{Introduction}
Observation of neutrinoless double beta decay ($0\nu\beta\beta$) would signify a ground-breaking discovery of beyond-the-Standard Model (BSM) by demonstrating the violation of lepton number conservation and confirming the Majorana nature of neutrinos. The canonical mode, in which two neutrons decay into two protons and two electrons without the emission of antineutrinos, has been a focus of extensive experimental and theoretical efforts. However, various extensions of the SM predict alternative modes of $0\nu\beta\beta$ that involve the emission of one or more light scalar bosons, often referred to as Majorons.

Originally introduced in the context of spontaneous lepton number violation, Majorons are massless (or very light) Nambu–Goldstone bosons that couple weakly to neutrinos and arise in many seesaw-inspired models of neutrino mass generation~\cite{Chikashige:1980qk,Chikashige:1980ui,Gelmini:1980re,deGiorgi:2023tvn} and can also play a role of dark matter, see, e.g., refs.~\cite{Berezinsky:1993fm,Garcia-Cely:2017oco,Brune:2018sab,Akita:2023qiz}. The process of neutrinoless double beta decay with Majoron emission involves two electrons and a Majoron emitted from the decaying nucleus, leading to a continuous energy spectrum that is distinct from both the $0\nu\beta\beta$ and the two-neutrino double beta decay ($2\nu\beta\beta$) modes~\cite{Doi:1987rx,Brune:2018sab,Blum:2018ljv}. Modifications to the ``ordinary'' Majoron process have been proposed in the form of, e.g., multi-Majoron models and charged Majorons~\cite{Burgess:1993xh,Bamert:1994hb,Hirsch:1995in}. More broadly, this idea can be generalised to a neutrinoless double beta decay with emission of any light scalar or pseudoscalar boson. This (pseudo-)scalar boson then either carries lepton number of two units, or the corresponding coupling involving the new boson violates lepton number by two units. Moreover, the scalar-neutrino coupling is intriguing also from the point of view of the neutrino self-interactions often discussed in various contexts~\cite{Blinov:2019gcj,DeGouvea:2019wpf,Berryman:2022hds,Fiorillo:2023ytr,Suliga:2024nng}.

Another often-discussed BSM (pseudo-)scalar is the axion, originally introduced to solve the strong CP problem~\cite{Peccei:1977hh,Peccei:1977ur,Weinberg:1977ma,Wilczek:1977pj} while simultaneously providing a dark matter candidate~\cite{Preskill:1982cy,Abbott:1982af,Dine:1982ah,Turner:1983he,Turner:1985si}. While axions and axion-like-particles typically have derivative couplings, in the presence of CP-violating new physics at large scales, CP-violating non-derivative axion interactions with the SM can be induced~\cite{Dekens:2022gha,Plakkot:2023pui,DiLuzio:2023cuk,DiLuzio:2023lmd}. Models where the axion (or axion-like-particle) also acts as the Majoron have been proposed~\cite{Ma:2017vdv,Cuesta:2021kca}.

The advancement of modern neutrinoless double beta decay and dark matter direct detection experiments has resulted in experiments with significantly larger masses and prolonged data collection periods~\cite{NEMO-3:2009fxe,NEMO-3:2016mvr,NEMO-3:2016qxo,nEXO:2018ylp,Arnold:2018tmo,CUPID:2019imh,KamLAND-Zen:2019imh,NEMO-3:2019gwo,legendcollaboration2021legend1000,XENON:2022evz}. Upcoming experiments will  record a large number of $2\nu\beta\beta$ events with minimal background interference, allowing for highly accurate measurements of the electron energy spectrum. Interpreting these measurements requires a careful description of the SM ($2\nu\beta\beta$) signal (see, e.g., refs~\cite{Simkovic:2018rdz,Nitescu:2024tvj,Morabit:2024sms} for recent developments), as well as possible BSM signals. 

This paper aims to investigate more generally the theoretical underpinnings of scalar-emitting \ovbb\;decay modes, which we call here \ovbbp, including the associated challenges stemming from the theoretical uncertainties on the double beta decay spectra. In particular, we want to improve the description of the \ovbbp\;spectrum compared to the seminal earlier papers \cite{Doi:1987rx,Brune:2018sab,Blum:2018ljv}. In ~\cref{sec:theory}, after introducing the effective low-energy couplings involving  (pseudo-)scalar bosons, we derive the corresponding rates for \ovbb\;with emission of this boson, while extending the state-of-the art framework for the canonical \ovbb\;decay without any additional particles in the final state. In doing so, we cover also the long-range \ovbb\;decay mechanisms involving sterile neutrinos. After that, in ~\cref{sec:stat}, we present the statistical procedure to be employed when deriving the limits on BSM physics from the \tvbb\;decay spectra, while paying special attention to theoretical uncertainties on these distributions. Employing this statistical analysis, we obtain the current experimental bounds on the effective scalar-neutrino couplings derived from double beta decay searches and explore the sensitivity of upcoming experiments in \cref{sec:results}. After discussing scalars heavier than the Q-value of the double beta decay under consideration in \cref{sec:offshell}, and the case of scalar emission via exotic right-handed lepton currents in \cref{sec:rhc}, we conclude in \cref{sec:conclusion}.

\section{Neutrinoless double beta decay with scalar emission}
\label{sec:theory}

We consider a general interaction of a BSM scalar to neutrinos of the form
\begin{align}
    \mathcal{L}\supset \sum_{i,j}\phi\bar{\nu}_i(g_{ij}+\gamma_5 \,\psg_{ij})\nu_j+\text{h.c.}\,,
    \label{eq:phinunu}
\end{align}
where $\nu_{i,j}$ denote Majorana neutrino mass eigenstates, with $i,j = 1,\,2,\,3$ for active neutrinos. Below, we also discuss the cases where sterile neutrinos are involved in the vertex, and connect the couplings $g_{ij}$ and $\psg_{ij}$ (and their sterile neutrino counterpart) to the effective coupling parameter $g_{ee}$. 

\subsection{Scalar-neutrino couplings and the corresponding matrix elements}

We are interested in computing the decay rate for $X \rightarrow Y + e^- +e ^- + \phi$ where $X$ and $Y$ are nuclear ground states with total angular momentum zero and positive parity (i.e., $\ket{0^+}$ states). Similar to \ovbb, the amplitude for this process can be computed as~\cite{Cirigliano:2017djv,Cirigliano_2018} 
\begin{align}
    \mathcal{A} = \bra{0^+}\sum_{m,n}\int \frac{d^3\vect q}{(2\pi)^3}e^{i\vect q \cdot\vect r}V(\vect q^2)\ket{0^+}\,,
    \label{eq:amp_pot}
\end{align}
where $\vect r = \vect r_n - \vect r_m$ for nucleons $m$ and $n$, and $\vect q$ is the momentum difference between the initial and final state nucleons. Here $|0^+\rangle$ and $\langle 0^+|$ states denote, respectively, the initial and final-state nucleus and $V(\vect q^2)$ denotes a neutrino `potential' that also contains the electron and scalar information. 

The amplitude can be written as 
\begin{align}
\mathcal{A} = \frac{(g_A^\text{eff})^2G_F^2V_{ud}^2m_e}{\pi\,R}\mathcal{A}^\phi\,\bar{u}(p_1)P_Ru^c(p_2)\,,
\label{eq:ampfactor}
\end{align}
for the Fermi constant $G_F$, the up-down CKM element $V_{ud}$, nuclear radius $R = 1.2 \,A^{1/3}$~fm with mass number $A$, the effective axial-vector coupling constant $g_A^\text{eff}$, and the electron momenta $p_{1,2}$, so that $\mathcal{A}^\phi$ depends only on the nuclear and hadronic matrix elements, allowing us to map it to the \ovbb\;matrix elements as discussed below. While other leptonic structures are possible (see \cref{sec:rhc}), the form of \cref{eq:ampfactor} is the most common one.

\begin{figure}[t!]
\centering
\begin{subfigure}[t]{0.45\textwidth}
\centering
\begin{tikzpicture}[scale=0.8, transform shape]
\begin{feynman}
\vertex (a); 
\node[blob,fill=black] (bl) at (a) ;
\vertex [above right=of a] (b);
\vertex [right=of b] (c);
\vertex [right=of c] (d);
\vertex [below right=1.2 of d] (e1);
\vertex [right=of d] (e);
\vertex [below right=of e] (f);
\node[blob,fill=black] (bl) at (f);
\vertex [below right=of a] (g) ;
\vertex [right=of g] (v2);
\vertex [right=of v2] (t2);
\vertex [above right=1.2 of t2] (e2);
\vertex [below left=of f] (h) ;
\vertex [above=1 of v2] (m1) ;
\vertex [right=0.7of m1] (m2) {\(\phi\)} ;
\diagram* {
(a) -- [quarter left, edge label = \(n\)] (b) -- [fermion] (e) -- [quarter left, edge label = \(p\)] (f),
(a) -- [quarter right, edge label'= \(n\)] (g) -- [fermion] (h) -- [quarter right, edge label' = \(p\)] (f),
(v2) -- [fermion] (e2),
(c) -- [fermion] (e1),
(v2) --  (c),
(m1) -- [scalar] (m2),
};
\vertex[right= 0.1 of e2]{\(e^-\)};
\vertex[right= 0.1 of e1]{\(e^-\)};
\vertex[blob,fill=gray,scale=0.5] at (c) {};
\vertex[blob,fill=gray,scale=0.5] at (v2) {};
\end{feynman}
\end{tikzpicture}
\end{subfigure}
\begin{subfigure}[t]{0.45\textwidth}
\centering
\begin{tikzpicture}[scale=0.8, transform shape]
\begin{feynman}
\vertex (a); 
\node[blob,fill=black] (bl) at (a) ;
\vertex [above right=of a] (b);
\vertex [right=of b] (c);
\vertex [right=of c] (d);
\vertex [below right=1.2 of d] (e1);
\vertex [right=of d] (e);
\vertex [below right=of e] (f);
\node[blob,fill=black] (bl) at (f);
\vertex [below right=of a] (g) ;
\vertex [right=of g] (v2);
\vertex [right=of v2] (t2);
\vertex [above right=1.2 of t2] (e2);
\vertex [below left=of f] (h) ;
\vertex [above=1 of v2] (m1) ;
\vertex [right=0.7of m1] (m2) {\(\phi\)} ;
\diagram* {
(a) -- [quarter left, edge label = \(n\)] (b) -- [fermion] (e) -- [quarter left, edge label = \(p\)] (f),
(a) -- [quarter right, edge label'= \(n\)] (g) -- [fermion] (h) -- [quarter right, edge label' = \(p\)] (f),
(v2) -- [fermion] (e1),
(c) -- [fermion] (e2),
(v2) --  (c),
(m1) -- [scalar] (m2),
};
\vertex[right= 0.1 of e2]{\(e^-\)};
\vertex[right= 0.1 of e1]{\(e^-\)};
\vertex[blob,fill=gray,scale=0.5] at (c) {};
\vertex[blob,fill=gray,scale=0.5] at (v2) {};
\end{feynman}
\end{tikzpicture}
\end{subfigure}
\caption{Leading order contributions to \ovbbp. The grey circles denote the effective weak vertex, and the black circles indicate the bound states of nucleons. A neutrino is exchanged between the nucleons, which radiates the scalar $\phi$.}
\label{fig:ovbb_feynman}
\end{figure}
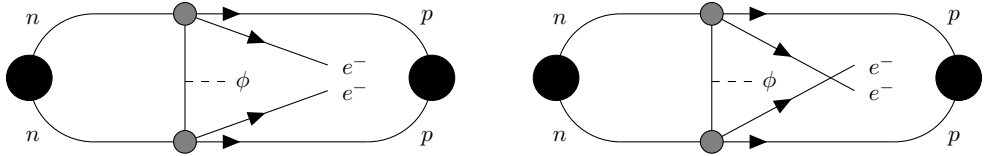

The neutrino potential $V(\vect q^2)$ is inserted between the initial and final $\ket{0^+}$ nuclear states, and is given from the diagrams \cref{fig:ovbb_feynman}. The first diagram computes to
\begin{align}
    V_1 = \;&-4i(\tau_1^+\tau_2^+)V_{ud}^2\frac{G_F^2}{2}J^\mu(1)J^\nu(2)\nonumber\\
    &\times \sum_{i,j}\mix_{ei}\,\mix_{ej}\bar{u}(p_1)\gamma_\mu P_L \frac{\slashed q_\nu +m_i}{(q_\nu^2 - m_i^2)} (g_{ij}+\gamma_5\,\psg_{ij})\frac{(\slashed q_\nu - \slashed{p}_\phi) + m_j}{(q_\nu-p_\phi)^2 - m_j^2})\gamma_\nu P_R \,u^c(p_2)\,,
\end{align}
where $J^\mu(n) = g_V\,v^\mu - 2g_A^\text{eff}\,S^\mu + \order{1/m_N}$ for the nucleon line $n$, with the nuclear vector coupling $g_V\approx1$, and the nucleon four-velocity and spin $v^\mu$ and $S^\mu$ respectively. The mixing angles of the $i$-th neutrino mass eigenstate with the electron neutrino are given by $\mathcal{U}_{ei}$, and $\tau^+ = \frac{\tau^1 + i \tau^2}{2}$ projects out the neutron to proton transition. The neutrino emitted from the top nucleon carries a momentum $q_\nu$. The neutrino masses are given by $m_{i,j}$ and the scalar carries away a momentum $p_\phi$.

We will consider the potential (long-distance) region of the neutrino momentum $q_\nu$, such that $q_\nu^0\ll|\vect q_\nu|\sim m_\pi$ can be ignored. We note however that the following discussion will follow through also for short-distance exchanges involving neutrino momenta in the hard region, i.e., with $q_\nu^0 \sim |\vect q_\nu|\sim\Lambda_\chi$ for the chiral scale $\Lambda_\chi \sim$~GeV~\cite{Cirigliano:2018hja}. The corrections from these short-range contributions have been shown to be sizeable~\cite{Dekens:2024hlz}, and we do include them in this work for phenomenological studies.

The ultrasoft region capturing virtual neutrinos with momenta comparable to the Q-value of the reactions cannot be described in the same way as for $0\nu\beta\beta$ because the momenta carried by the outgoing scalar can no longer be neglected. However, the ultrasoft region is a next-to-next-to-leading order effect\footnote{In specific scenarios where all sterile neutrinos are light compared to nuclear scales and neutrino masses are arising from the type-I seesaw, the leading-order amplitude vanishes and ultrasoft contributions become relevant \cite{Dekens:2023iyc}. We will not discuss this scenario here.} \cite{Cirigliano:2017tvr} as explicitly confirmed in refs.~\cite{Dekens:2024hlz,Castillo:2024jfj} and we will neglect it here.

In the potential and hard regimes we can neglect the electron and scalar four-momentum given that the Q-value of the decay is at most a few MeVs. With these approximations, the potential obtained by adding the two diagrams in \cref{fig:ovbb_feynman} can be written as 
\begin{align}
     V(\vect q^2) = \,&4i(\tau_1^+\tau_2^+)V_{ud}^2G_F^2\sum_{i,j}\mix_{ei}\,\mix_{ej}\Bigg[\frac{\vect q^2}{(\vect q^2 + m_i^2)(\vect q^2 + m_j^2)}(g_{ij} + \psg_{ij} ) \nonumber\\&- (g_{ij} - \psg_{ij})\frac{m_i\,m_j}{(\vect q^2 + m_i^2)(\vect q^2 + m_j^2)}\Bigg]\bar{u}(p_1) P_Ru^c(p_2)\, J(1)\cdot J(2)\,,
    \label{eq:leptonline}
\end{align}
which can then be inserted in \cref{eq:amp_pot} to compute the amplitude.
The form of the leptonic structure now justifies the factorisation in \cref{eq:ampfactor}. We will now study this general potential in several limiting cases.

\subsubsection{Scalar coupling to active neutrinos} In case of scalar coupling to only active neutrinos, $m_i\ll |\vect{q}|$, and we can thus also ignore the second term, as well as the masses in the neutrino propagators, in \cref{eq:leptonline}. This gives
\begin{align}
    V(\vect q^2) = 4i(\tau_1^+\tau_2^+V_{ud}^2 G_F^2)\sum_{i,j}U_{ei}\,U_{ej}\left(g_{ij}+\psg_{ij}\right)\frac{1}{\vect q^2}\bar{u}(p_1) P_Ru^c(p_2)\,J(1)\cdot J(2)\,,
\end{align}
where $U_{ek}$ are the PMNS elements. This is similar to the regular \ovbb\;potential, where an additional overall factor of neutrino mass appears. The hadronic part of the matrix element in both cases is the same, and thus, as noted in ref.~\cite{doi:1985dx}, in this scenario we can directly use the \ovbb\;nuclear matrix elements (NMEs) for the massless neutrino limit, as discussed in refs.~\cite{Dekens:2023iyc,Dekens:2024hlz,Cirigliano:2024ccq}, i.e., $\mathcal{A}^\phi = \amp(0)$. The \ovbb\;amplitudes used are described in \cref{app:ovbbamp}. Further, we can identify the overall coupling, that will allow us to write the half-life in a~factorised form as shown in \cref{eq:majHL}, as $g_{ee} = \sum_{i,j}U_{ei}U_{ej}(g_{ij}+\psg_{ij})$. 

\subsubsection{Scalar coupling to sterile neutrinos} If the scalar emission takes place through its coupling to massive BSM neutrinos, however, the second term in \cref{eq:leptonline} cannot be neglected. Consider the case of a scalar interacting with a single species of massive BSM neutrinos, such that
\begin{align}
    \mathcal{L} \supset \phi\bar{N}(\stc +\gamma_5\,\stcg)N+\text{h.c.}\,.
\end{align} 
Using $\Theta_{eN}$ to denote the mixing of the SM singlet sterile neutrino with the electron flavour neutrino, we get 
\begin{align}
    V(\vect q^2) = \,&4i(\tau_1^+ \tau_2^+)\Theta_{eN}^2\left[\stcg\frac{1}{\vect q^2 + M^2} + \stc\frac{q^2 - M^2}{(\vect q^2 + M^2)^2}\right]\bar{u}(p_1) P_R u^c(p_2)\,J(1)\cdot J(2)\nonumber\\
    =\,&4i(\tau_1^+ \tau_2^+)\Theta_{eN}^2\,\bar{u}(p_1)P_R u^c(p_2)\,J(1)\cdot J(2)\nonumber\\&\times\left[\stcg\frac{1}{\vect q^2 + M^2} + \stc\left(\frac{1}{\vect q^2 + M^2}-\frac{2M^2}{(\vect q^2 + M^2)^2}\right)\right]\,,
    \label{eq:sterileamp}
\end{align}
where $M$ is the mass of the sterile neutrino. The first term reduces to the \ovbb\;matrix element with a dependence on neutrino mass, and we can use the parametrised form $\mathcal{A}^\phi(M) = \amp(M)$ described in refs.~\cite{Dekens:2024hlz,Cirigliano:2024ccq}. The second term is a bit more involved. Nonetheless, as shown in \cref{app:ovbbamp}, all the mass-dependence of the potential is explicitly given in \cref{eq:sterileamp}. Thus, using $\frac{\partial}{\partial M} \left(\frac{1}{\vect q^2+M^2}\right) = -\frac{2M}{\left(\vect q^2 + M^2\right)^2}$, we can approximate the amplitude without computing the NMEs explicitly as
\begin{align}
    \mathcal{A}^\phi(M) = \left(1 + M\frac{\partial}{\partial M}\right)\amp(M).
    \label{eq:sterileampder}
\end{align}
Note that the parameters in the interpolation formulae used to describe the mass-dependent NMEs (see \cref{eq:gnu_int}) are set such that the amplitude is continuous at the interface $M=2$~GeV, beyond which the neutrino is integrated out at the quark level and the amplitude is computed as a dimension-9 operator. This means that the use of \cref{eq:sterileampder} requires a~modification to the amplitude, which we do by refitting the interpolation parameters to ensure smoothness at the boundary. Separately, we can identify the effective coupling for the scalar and pseudoscalar interactions as $g_{ee} = \Theta_{eN}^2\overset{\textbf{\fontsize{2pt}{2pt}\selectfont(--)}}{g}_{\phi N N}$.

\subsubsection{Mixed coupling} Finally, we also consider a scenario where the scalar may couple to one active and one sterile neutrino, i.e.,
\begin{align}
    \mathcal{L}\supset\sum_{i}\phi\bar{\nu}_i(g_{\phi N \nu_i}+\gamma_5 \, \bar{g}_{\phi N \nu_i})N+\text{h.c.}\,
\end{align}
Assuming $|\vect q|^2 \gg m_i \,M$, which should be true for sterile neutrino masses up to $\sim 10^8$~GeV, we can once again ignore the second term in \cref{eq:leptonline}.\footnote{For sterile masses beyond $\sim10^8$~GeV, the second term in \cref{eq:leptonline} becomes $\propto \frac{m_i}{M}\frac{1}{\vect q^2}$, and can be computed like the usual light neutrino \ovbb\;potential. We however restrict ourselves to masses much smaller than this.}  Ignoring the light neutrino mass in the propagator, we get $\mathcal{A}^\phi(M) = \amp(M)$ to be the corresponding NME, and the effective coupling is $g_{ee} = \Theta_{eN}\sum_iU_{ei}(g_{\phi N \nu_i} + \bar g_{\phi N \nu_i})$.

\subsection{The phase space factor}
\label{sec:PSF}

The phase space factor for the scalar-emitting process, for all the scenarios discussed above and assuming the scalar mass is below the Q-value of the isotope such that it is produced on-shell, is~\cite{Kotila:2015ata, EXO-200:2014vam}
\begin{align}
        \Omega^{\phi} &= \frac{1}{\ln{2}}\frac{(G_F V_{ud} g_A^{\text{eff}})^4}{128 \pi^7 R^2} \int_{m_e}^{E_i-E_f-m_e-m_\phi} dE_{e_1} F(Z_f,E_{e_1})|\vect p_{e_1}| E_{e_1} \nonumber\\
        & \times\int_{m_e}^{E_i-E_f-E_{e_1} - m_\phi} dE_{e_2} F(Z_f,E_{e_2})|\vect p_{e_2}| E_{e_2}\int dE_{\phi}|\vect p_{\phi}|\,\delta (E_i-E_f-E_{\phi}-E_{e_1}-E_{e_2})\,.
        \label{eq:majoronPSF}
\end{align}
The energies of the initial and final states are related to the Q-value as $E_i - E_f = Q + 2m_e$. $Z_f$ is the atomic number of the daughter nucleus, $E_{e_i},\,\vect p_{e_i}$ are the energies and three-momenta of the outgoing electrons, and $E_\phi\,, \vect p_\phi$ are the scalar energy and momentum, respectively. $F(Z,E_e)$ is the Fermi function accounting for the Coulomb interaction of the outgoing electrons in the nuclear field. For $0^+ \rightarrow 0^+$ transitions, the Fermi function consists of the radial components of the electron wavefunction $g_{-1}(E_e,r)$ and $f_1(E_e,r)$~\cite{Nitescu:2021},
\begin{align}
    F(Z, E_e) &= g_{-1}^2(E_e,R) + f_1^2(E_e,R)\,,
    \label{eq:fermifunction}
\end{align}
evaluated at the nuclear radius.

We adopt the approximation of the relativistic electron wavefunction in a uniform charge distribution in the daughter nucleus. In this approximation, the Fermi function is 
\begin{align}
    F(Z, E_e) &= \left[\frac{\Gamma(3)}{\Gamma(1)\Gamma(1+2\gamma_0)}\right]^2\left(2|\vect p_e| R\right)^{2(\gamma_0-1)}e^{\pi y}\left|\Gamma(\gamma_0 + i y)\right|^2\,,
    \label{eq:fermifunction_approx}
\end{align}
where $\gamma_0 = \sqrt{1 - (\alpha\,Z)^2}$, and $y = \alpha Z \frac{E_e}{|\vect p_e|}$.
A better description is obtained by including finite nuclear size and screening \cite{Nitescu:2021,Stefanik:2015twa}, radiative and atomic exchange corrections \cite{Nitescu:2024tvj}, and with the use of codes such as \texttt{RADIAL}~\cite{2019CoPhC.240..165S} as done in ref.~\cite{Kotila:2012zza}, or by using more precise analytical forms $g_{-1}(E_e)$ and $f_1(E_e)$ as done in, e.g., \cite{Nitescu:2021}. See ref.~\cite{Stefanik:2015twa} for a comparison of various methods.

A change of variables $\epsilon = E_{e_1} + E_{e_2} - 2 m_e$, $\Delta = E_{e_1} - E_{e_2}$ allows us to express the phase space integral in terms of the summed electron energy $\epsilon$ and their difference $\Delta$, which is useful given that summed energy is a primary observable in double beta decay experiments. 

\begin{figure}[h!]
    \centering
    \begin{subfigure}[b]{0.49\textwidth}
        \includegraphics[width=\linewidth]{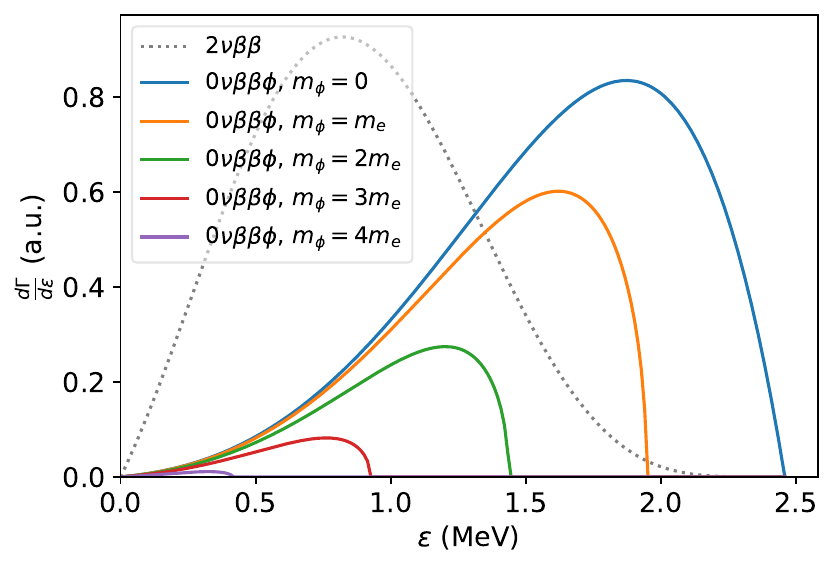}
    \end{subfigure}
    \begin{subfigure}[b]{0.49\textwidth}
        \includegraphics[width=\linewidth]{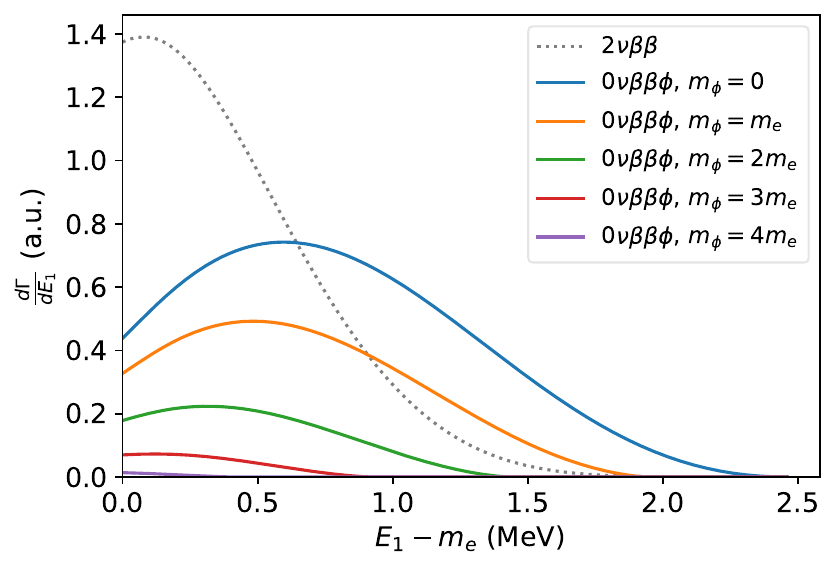}
    \end{subfigure}
    \caption{The summed electron energy (left) and single electron energy (right) spectra for \tvbb\;and \ovbb\;with scalar emission for various scalar masses for \Xe. The \tvbb\;and $m_\phi =0$ distributions are normalised to unity, while the rest are normalised to the $m_\phi=0$ distribution.}
    \label{fig:spectra}
\end{figure}

At the order we are working at, i.e., with the approximations we have made above, the half-life of a scalar-emitting neutrinoless double beta decay process can be given, using the phase space factor, the effective coupling, and the matrix element, in the form
\begin{align}
    \left(\thm\right)^{-1} = \frac{\Gamma^\phi}{\ln2} = \psfm |g_{ee}|^2 |\mathcal{A}^\phi|^2\,.
    \label{eq:majHL}
\end{align}
As discussed above, the NME of the process, $\mathcal{A}^\phi$, can in general depend on the masses of the neutrinos being exchanged. The phase space factor \psfm\;depends on the scalar mass. The model-dependent effective scalar coupling to the electron neutrino, $g_{ee}$, is a combination of the couplings $g_{ij}$, $\bar g_{ij}$ in \cref{eq:phinunu} (or their sterile neutrino counterparts), and the neutrino mixing angles.

\subsection{Angular correlation}
\label{sec:angular}

Apart from the energy distributions, another potentially useful observable is the angular correlation between the two outgoing electrons, particularly in light of experiments that have the ability to resolve individual final state electrons~\cite{NEMO:2006smm,SuperNEMO:2010wnd}. Following ref.~\cite{Kotila:2015ata}, the corresponding \ovbbp\;angular distribution of emitted electrons is 
\begin{align}
    \frac{d\Gamma^\phi}{d(\cos{\theta})} = \frac{1}{2}\Gamma^\phi(1+K^\phi \cos{\theta}),
    \label{eq:angcorrrate}
\end{align}
where $\theta$ is the angle between the emitted electrons and
\begin{align}
    K^\phi = -\frac{\Lambda^\phi}{\Gamma^\phi}
\end{align}
is the angular correlation factor. $\Lambda^\phi$ takes the form similar to $\Gamma^\phi$~\cite{Kotila:2015ata}, based on \cref{eq:majHL} and \cref{eq:majoronPSF},
\begin{align}
    \Lambda^\phi = \;&\frac{(G_F V_{ud} g_A^{\text{eff}})^4}{128 \pi^7 R^2}|g_{ee}|^2\int_{m_e}^{E_i-E_f-m_e-m_\phi} dE_{e_1}\int_{m_e}^{E_i-E_f-E_{e_1} - m_\phi} dE_{e_2}\int dE_{\phi}\\
        &\times E(Z_f,E_{e_1})E(Z_f,E_{e_2}) E_{e_1} |\vect p_{e_1}| E_{e_2} |\vect p_{e_2}| |\vect p_{\phi}|\,\delta (E_i-E_f-E_{\phi}-E_{e_1}-E_{e_2})|\mathcal{B}^\phi|^2\,.\nonumber
        \label{eq:angular}
\end{align}
With the leading-order approximation $\mathcal{B^\phi} = \mathcal{A}^\phi$ (see \cref{app:angular}), $K^\phi$ becomes independent of the nuclear structure, and then $\Lambda^\phi$ differs from $\Gamma^\phi$ only in replacing the Fermi function~\eqref{eq:fermifunction} by the function~\cite{Kotila:2015ata, Kotila:2012zza, Nitescu:2021}
\begin{align}
    E(Z,E_e) = 2g_{-1}(E_e)f_1(E_e)\,.
\end{align}
Using the form of the Fermi function shown in \cref{eq:fermifunction_approx}, we employ the approximation~\cite{Deppisch:2020mxv}
\begin{align}
    E(Z,E_e) \simeq \frac{|\vect p_e|}{E_e}F(Z,E_e)\,.
    \label{eq:fermi_mod}
\end{align}
\begin{figure}[h!]
    \centering
    \includegraphics[width=0.6\textwidth]{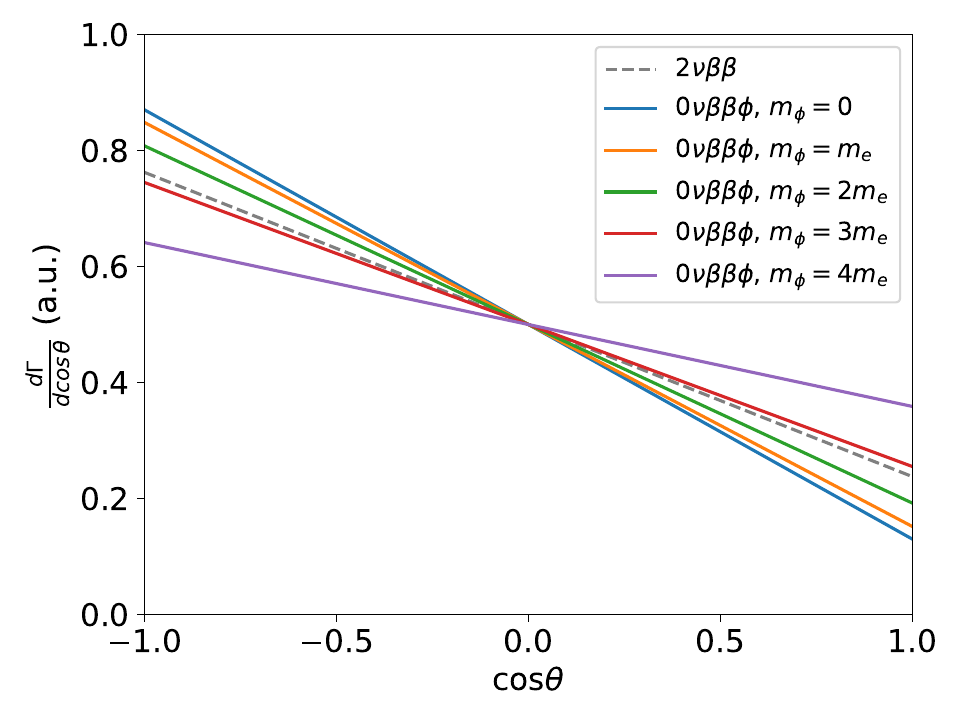}
    \caption{The angular correlation for \tvbb\;and \ovbb\;with scalar emission for various scalar masses for \Xe. All the distributions are normalised to unity.}
    \label{fig:spectra_angular}
\end{figure}

\Cref{fig:spectra_angular} shows the angular correlation for the \tvbb\;and \ovbbp\;spectra for various scalar masses. We see that for small $m_\phi$, a back-to-back emission of electrons is preferred, similar to \tvbb, but for larger masses, this preference vanishes. As the slope decreases monotonically with the mass, for some intermediate value of $m_\phi$, the slope becomes similar to \tvbb, making the signal harder to distinguish given the lack of shape distortion. As such, for a particular range of masses, a study of the angular correlation alone will not suffice to distinguish new physics from the \tvbb\;background, as can be seen explicitly in \cref{fig:angcorr}.

\section{Statistical analysis and theoretical uncertainties}
\label{sec:stat}

\subsection{Double beta decay spectrum \& uncertainties}
\label{sec:2nubb}

The major SM background to \ovbbp\;that we consider here is \tvbb\;decays, and we follow the description given in recent developments in the theoretical prediction of the decay rates~\cite{Simkovic:2018rdz,Morabit:2024sms}, including higher-order chiral corrections, to model this background as accurately as possible.
The \tvbb\;decay rate is given by~\cite{Morabit:2024sms}
\begin{align}
      \frac{\Gamma^{2\nu}}{\ln 2} &\simeq (g_A^{\text{eff}})^4|\mathcal{M}_{GT}^{(-1)}|^2\Delta_0\left[G^{2\nu}_0 + \xi_{31}\frac{\Delta_2}{\Delta_0}G^{2\nu}_2 + \frac{1}{3}\xi_{31}^2 G^{2\nu}_{22} + \left(\frac{1}{3}\xi_{31}^2 + \xi_{51}\frac{\Delta_2}{\Delta_0}\right)G^{2\nu}_4 + \frac{G^{2\nu}_M}{\Delta_0}\right]\,,
      \label{eq:2nugamma}
\end{align}
where the phase space factors $G^{2\nu}_i$ are
\begin{align}
    G^{2\nu}_i =\; &\frac{1}{\ln 2}\frac{(G_F\,V_{ud})^4}{8\pi^7m_e^2}\int_{m_e}^{E_i - E_f - m_e} dE_{e_1} \int_{m_e}^{E_i - E_f - E_{e_1}} dE_{e_2} \int_0^{E_i - E_f - E_{e_1} - E_{e_2}} dE_{\nu_1}\int dE_{\nu_2} \nonumber\\
    &\times F(Z_f, E_{e_1})F(Z_f, E_{e_2})E_{e_1}|\vect p_{e_1}|E_{e_2}|\vect p_{e_2}|E_{\nu_1}^2E_{\nu_2}^2  \mathcal{A}^{2\nu}_i\nonumber\\
    &\times\delta(E_{\nu_2} - E_i + E_f + E_{e_1} + E_{e_2} + E_{\nu_1})\,,
    \label{eq:2nupsf}
\end{align}
with 
\begin{align}
    \mathcal{A}^{2\nu}_0&=1,\quad \mathcal{A}^{2\nu}_2=\frac{\varepsilon_K^2 + \varepsilon_L^2}{(2m_e)^2},\quad \mathcal{A}^{2\nu}_{22}=\frac{\varepsilon_K^2 \varepsilon_L^2}{(2m_e)^4},\quad \mathcal{A}^{2\nu}_4 = \frac{\varepsilon_K^4 + \varepsilon_L^4}{(2m_e)^4}\nonumber\\
    \mathcal{A}^{2\nu}_M&=\frac{2g_M}{3m_N g_A^{\text{eff}}}\frac{(E_{e1} + E_{e2})(2E_{e1}E_{e2}-m_e^2)}{E_{e1}E_{e2}}\,,
\end{align}
being obtained by expanding the matrix elements in the lepton energy factors
\begin{align}
    \epsilon_K = (E_{e_2} + E_{\nu_2}-E_{e_1} - E_{\nu_1})/2\,,\quad \epsilon_L = (E_{e_1} + E_{\nu_2}-E_{e_2} - E_{\nu_1})/2\,.
\end{align}

The $\xi_{i1}$ parameters are the ratio of leading order pieces of the Gamow-Teller matrix element to the higher order ones, $\xi_{i1}=\mathcal{M}_{GT}^{(-i)}/\mathcal{M}_{GT}^{(-1)}$, where the Gamow-Teller pieces are defined as~\cite{Morabit:2024sms}
\begin{align}
    M_{GT}^{(-2m-1)} = m_e\left(2m_e\right)^{2m}\sum_n \frac{\bra f \sum_k \boldsymbol{\sigma}_k\tau^+_k \ket n \cdot\bra n \sum_l \boldsymbol{\sigma}_l\tau^+_l\ket i}{\left(E_n - \frac12(E_i + E_f)\right)^{(2m+1)}}\,,
\end{align} 
in terms of the transition matrix elements for the initial, final, and intermediate states $\ket i,\,\ket f,\,\ket n$ respectively, and where $k,\,l$ denote the nucleons inside the nucleus. The parameters $\Delta_0$ and $\Delta_2$ encode higher-order chiral corrections to the decay rate arising from both weak magnetic and double-weak two-nucleon pion-exchange currents. $g_M = 1+\kappa_1 \simeq 4.7$ is the nucleon magnetic coupling, with $m_N$ being the nucleon mass. Several of these parameters come with considerable uncertainties, the effect of which are discussed in \cref{sec:results}. \Cref{tab:params} gives the central values of the parameters that we use in this work, unless specified otherwise. We use nuclear shell model (NSM) determinations of the matrix elements and parameters wherever possible, and QRPA results where NSM results are not available, e.g., for $^{82}$Se and $^{100}$Mo. Among these parameters, $\xi_{31}$ is the biggest source of uncertainty when it comes to determining limits on the scalar coupling; see \cref{fig:xi31unc}, where we use the range $\xi_{31}\in[0.12,0.20]$, and the accompanying discussion.

\begin{table}[ht!]
    \centering
    \begin{tabular}{|c|c|c|c|c|}
    \hline
      & \Xe & \Ge & $^{82}$Se & $^{100}$Mo \\
      \hline
     $\xi_{31}$ & 0.16 & 0.12 & 0.13 & 0.56 \\
     $\xi_{51}$ & 0.042 & 0.022 & 0.027 & 0.21 \\
     $\Delta_0$ & 1.077 & 1.020 & 1.038 & 1.013 \\
     $\Delta_2$ & 1.043 & 1.014 & 1.024 & 1.012 \\
     $M_{GT}^{(-1)}$ & 0.013 & 0.058 & 0.050 & 0.11 \\ 
     \hline
     $Q$ & 2.458 & 2.039 & 2.995 & 3.034 \\
     \hline
    \end{tabular}
    \caption{Central values of \tvbb\;parameters, and \tvbb\;Q-values, used in this work~\cite{Morabit:2024sms,Simkovic:2018rdz,Menendez:2017fdf,Hyvarinen:2015bda}. The interpolation formulae from ref.~\cite{Dekens:2020ttz} have been employed to compute $\Delta_{0,2}$ for $^{100}$Mo and $^{82}$Se.}
    \label{tab:params}
\end{table}

The angular distribution of the electrons can be obtained as described for the \ovbbp\; decay in \cref{sec:angular}, i.e.,
 \begin{align}
     \frac{d\Gamma^{2\nu}}{d(\cos\theta)} = \frac12\Gamma^{2\nu}\left(1 -\frac{\Lambda^{2\nu}}{\Gamma^{2\nu}}\cos\theta\right)\,,
 \end{align}
 where $\Lambda^{2\nu}$ can be obtained by the replacements~\cite{Nitescu:2021}
 \begin{align}
     \frac13\xi_{31}^2G_{22}^{2\nu}\rightarrow\frac59\xi_{31}^2\overline{G}_{22}^{2\nu},\qquad \left(\frac13\xi_{31}^2 + \xi_{51}\frac{\Delta_2}{\Delta_0}\right)G_4^{2\nu}\rightarrow\left(\frac29\xi_{31}^2 + \xi_{51}\right)\overline G_4^{2\nu}\,,
 \end{align}
in \cref{eq:2nugamma}, where the modified phase space factors $\overline G_i^{2\nu}$ contain the functions \cref{eq:fermi_mod} instead of the Fermi functions. Since they have not yet been computed explicitly, we will ignore the correction due to the magnetic piece $\bar G^{2\nu}_M$, and set $\Delta_2 = \Delta_0 = 1$, in $\Lambda^{2\nu}$.

In addition to the \tvbb\;parameters mentioned above, another source of uncertainty that potentially affects our analysis is the use of the Fermi function given in \cref{eq:fermifunction_approx}. In principle, for a complete analysis, one should compute the electron wavefunctions numerically, considering effects such as screening. Based on the comparison in ref.~\cite{Stefanik:2015twa}, the most affected quantity will be $\Lambda^{2\nu,\phi}$, and consequently the angular correlations. $F(Z,E_e)$ is dominated by the $g_{-1}(E_e)$ component and does not vary significantly across different methods. The difference in the spectra due to these corrections will be
minimal in the scenarios we consider; for example, our computed value of $\Omega_\phi/g_A^4$ with massless scalar is $\sim 2.7\cdot10^{-16}~\text{yr}^{-1}$, compared to the values $\sim 4.1\cdot10^{-16}\text{yr}^{-1}$ and $\sim 4.3\cdot10^{-16}\text{yr}^{-1}$ in refs.~\cite{Kotila:2015ata,EXO-200:2014vam} where exact Dirac wavefunctions are used, and the effects of finite nuclear size and
electron screening are also considered. Their effect on the sensitivity is thus small in the presence of NME uncertainties discussed below, and we use the simple analytical form of \cref{eq:fermifunction_approx}.

\subsection{Effect of uncertainties in the NMEs}

The \ovbbp\;NMEs involved in the prediction of the decay rate also comes with several possible sources of uncertainties. For neutrino masses below 2 GeV, the contribution from the long-range (potential) and short-range (hard)~\cite{Cirigliano:2018hja} exchange of neutrinos is taken into account. The ultrasoft contribution described in refs.~\cite{Dekens:2023iyc,Dekens:2024hlz} is not considered here, as the effects are subleading \cite{Castillo:2024jfj}.
The potential and hard contributions consist of NMEs and low energy constants (LECs), which need to be computed using, e.g., nuclear many-body methods, lattice QCD, or interpreted from experiments. Here, we use values computed within NSM, and use interpolation formulae as described in refs.~\cite{Dekens:2024hlz,Cirigliano:2024ccq}. The NME so obtained can have uncertainties arising from, among others, 1) the many-body method chosen, e.g., NSM vs. QRPA, 2) theoretical uncertainties within the particular many-body computation, 3) uncertainties introduced while translating the computed values to the interpolation formula showing up, e.g., in the form of imperfections in the fit parameters.

For neutrino masses above 2 GeV, the neutrino is integrated out at the quark level, and thus the amplitude is obtained from a dim-9 operator involving four quarks and two electrons. The NME so obtained contains, on top of short-distance NMEs, the LECs $g^{NN}_1,\,g_1^{\pi N},\,g_1^{\pi\pi}$ which  lead to additional uncertainties. We will here  follow the recipe in ref.~\cite{Dekens:2024hlz}, and use $g_1^{NN} = (1+3g_A^2)/4$, $g_1^{\pi N}=0$, and $g_1^{\pi\pi} = 0.36$~\cite{Nicholson:2018mwc}.

The effect of the uncertainties in NMEs on the obtained limits to the coupling of BSM scalars to active neutrinos is briefly discussed in \cref{sec:results}. The amount of uncertainty in the limits for other couplings is expected to be similar. However, for $\stc$, the location where the amplitude crosses zero critically depends on the exact values of amplitudes across the neutrino mass scale, and thus the effect is hugely amplified in the form of a region where one might not be able to set any limits on the coupling. This scenario is discussed in more detail in \cref{sec:results}, and visualised in \cref{fig:sterile_limit}.

\subsection{Statistical analysis}

With the updated phase space factors and amplitudes at hand, we focus on subtle deviations from the SM prediction originating from \ovbbp, and we put limits on the strength of the BSM signal quantified by the relevant coupling constant. To determine its maximum value such that the signal is still indistinguishable from the SM \tvbb, we employ the standard frequentist approach outlined in ref.~\cite[sec.~5]{Bolton:2020ncv}. We briefly describe the procedure below.

Suppose that a given experiment measures an observable $T$ with a resolution $\Delta T$. This observable can be summed electron energy $\epsilon$, single electron energy $E_{e_{1,2}}$, or angle between the emitted electrons, $\cos\theta$. When comparing \cref{fig:spectra} and \cref{fig:spectra_angular}, one can see qualitatively distinct signatures in different channels. Consequently, certain observables may be more suitable for probing specific BSM scenarios. In the following, we therefore compare the analysis for all the aforementioned observables.

Let $N_\text{tot}$ denote the total number of events in a given data set $\mathcal{D}$, distributed over a~number of bins $N_\text{bins}$. Assuming that the events follow a known differential distribution composed of the SM background (the source of which is mainly \tvbb), and the BSM signal in the form of \ovbbp, the expected number of events per $i$-th bin is
\begin{align}
    N_\text{exp}^{(i)} = N_\text{sig}^{(i)} + N_\text{bkg}^{(i)} = \frac{N_\text{tot}}{\Gamma^{2\nu} + \Gamma^\phi}\int_{T_i}^{T_{i+1}} dT\,\left(\frac{d\Gamma^\phi}{dT}+\frac{d\Gamma^{2\nu}}{dT}\right)\,,
    \label{eq:1D-dist}
\end{align}
where $N_\text{sig}^{(i)}$ and $N_\text{bkg}^{(i)}$ are the numbers of signal and background events per bin, respectively. 
The corresponding decay rates $\Gamma^\phi$ and $\Gamma^{2\nu}$ are defined in \cref{eq:majHL} and \cref{eq:2nugamma}. $T_{i,i+1}$ denote the $i$-th bin edges.

Since the signal is normalised to the total decay rate, the analysis essentially looks for changes in the shape of the spectra from what is expected, in a manner similar to the analysis in ref.~\cite{Morabit:2024sms}. Note that this means the uncertainties in overall factors such as $g_A^\text{eff}$ and $M_{GT}^{(-1)}$ cancel out during the normalisation. At leading order, effectively the uncertainty comes from the ratio of overall NMEs of \ovbbp\;and \tvbb\;decays (which can be parametrised as an uncertainty in just the \ovbbp\;NME as shown in \cref{fig:active_limit}), and from the $\xi_{i1}$ and $\Delta_{0,2}$ parameters at higher orders.

To test the outlined hypothesis, i.e., that we expect the data to contain a signal originating from massive scalar with parameters $\param^\phi =(|g_{ee}|,m_\phi)$, we use the test statistic
\begin{align}
    q_{\param^\phi} = -2\ln{\frac{\mathcal{L}(\mathcal{D}|\param^\phi,\hat{\hat{\param}}^{2\nu})}{\mathcal{L}(\mathcal{D}|\hat{\param}^\phi,\hat{\param}^{2\nu})}}\,,
    \label{eq:test_statistic}
\end{align}
where $\mathcal{L}(\mathcal{D}|\param^\phi,\hat{\hat{\param}}^{2\nu})$ and $\mathcal{L}(\mathcal{D}|\hat{\param}^\phi,\hat{\param}^{2\nu})$ are likelihood functions of observing data $\mathcal{D}$ given the null and alternative hypotheses, respectively. Here, $\param^{2\nu}=\{g_A^{\text{eff}},\mathcal{M}_{GT}^{(-1)},\xi_{i1},\Delta_0,\Delta_2\}$ represents the set of parameters associated with \tvbb, and the quantities with hats denote the values of the parameters that maximise $\mathcal{L}$,
\begin{align}
    \mathcal{L}(\mathcal{D}|\param^\phi,\hat{\hat{\param}}^{2\nu}) &= \max_{\param^{2\nu}}{\mathcal{L}(\mathcal{D}|\param^\phi,\param^{2\nu})}\,,\\
    \mathcal{L}(\mathcal{D}|\hat{\param}^\phi,\hat{\param}^{2\nu}) &= \max_{\param^\phi,\param^{2\nu}}{\mathcal{L}(\mathcal{D}|\param^\phi,\param^{2\nu})} \,.
\end{align}
In other words, $\hat{\param}^\phi$ and $\hat{\param}^{2\nu}$ are the maximum-likelihood (ML) estimators of $\param^\phi$ and $\param^{2\nu}$, respectively, while $\hat{\hat{\param}}^{2\nu}$ is the conditional ML estimator of $\param^{2\nu}$ for fixed $\param^\phi$. Thus, $\hat{\hat{\param}}^{2\nu}$ is a~function of $\param^\phi$.

$\mathcal{L}(\mathcal{D}|\param^\phi,\param^{2\nu})$ is constructed as the product of the Poisson probabilities $P(N_\text{obs}^{(i)}|N_\text{exp}^{(i)})$ over all bins, where $N_\text{obs}^{(i)}$ is the number of events per bin observed by the experiment. Since the test statistic \eqref{eq:test_statistic} involves the logarithm of the likelihood ratio -- equivalent to the difference of the logarithms -- instead of $\mathcal{L}(\mathcal{D}|\param^\phi,\param^{2\nu})$ we express the definition in terms of the log-likelihood function
\begin{align}
    -2\ln\mathcal{L}(\mathcal{D}|\param^\phi,\param^{2\nu}) = 2\sum_i^{N_\text{bins}}\left[N_{\text{exp}}^{(i)}(\param^\phi,\param^{2\nu}) - N_\text{obs}^{(i)} + N_\text{obs}^{(i)}\ln\left(\frac{N_\text{obs}^{(i)}}{N_\text{exp}^{(i)}(\param^\phi,\param^{2\nu})}\right)\right]\,.
\end{align}

In practice, the observed data $\mathcal{D}$ can either be real experimental measurements or simulated data sets from Monte Carlo simulations of the experiment. Instead, we use a representative artificial data set $\mathcal{D}_A$ called Asimov data set~\cite{Cowan:2010js}, for which the observed number of events in each bin is taken to be equal to the expected number of background events, $N_\text{obs}^{(i)}=N_\text{exp}^{(i)}(|g_{ee}|=0, \param^{2\nu}_{\text{obs}})$. In this case, the data are best described by the background-only hypothesis, i.e., the likelihood function is maximized when the signal strength is set to zero and $\hat{\param}^{2\nu}=\param^{2\nu}_{\text{obs}}$, 
\begin{align}
    -2\ln\mathcal{L}(\mathcal{D}_A|\hat{\param}^\phi,\hat{\param}^{2\nu}) &= 2\min_{\param^\phi,\param^{2\nu}} \sum_i^{N_\text{bins}}\Bigg[N_{\text{exp}}^{(i)}(\param^\phi,\param^{2\nu}) - N_\text{obs}^{(i)}(|g_{ee}|=0, \param^{2\nu}_{\text{obs}}) \nonumber\\
    &+ N_\text{obs}^{(i)}(|g_{ee}|=0, \param^{2\nu}_{\text{obs}})\ln\left(\frac{N_\text{obs}^{(i)}(|g_{ee}|=0, \param^{2\nu}_{\text{obs}})}{N_\text{exp}^{(i)}(\param^\phi,\param^{2\nu})}\right)\Bigg]=0\,.
\end{align}

We are interested in determining the strongest BSM signal that remains consistent with the data at a given confidence level. We need to solve the following equation
\begin{align}
    q_{\param^\phi} &=-2\ln\mathcal{L}(\mathcal{D}_A|\param^\phi,\hat{\hat{\param}}^{2\nu}) = 2\min_{\param^{2\nu}} \sum_i^{N_\text{bins}}\Bigg[N_{\text{exp}}^{(i)}(\param^\phi,\param^{2\nu}) - N_\text{obs}^{(i)}(|g_{ee}|=0, \param^{2\nu}_{\text{obs}}) \nonumber\\
    &+ N_\text{obs}^{(i)}(|g_{ee}|=0, \param^{2\nu}_{\text{obs}})\ln\left(\frac{N_\text{obs}^{(i)}(|g_{ee}|=0, \param^{2\nu}_{\text{obs}})}{N_\text{exp}^{(i)}(\param^\phi,\param^{2\nu})}\right)\Bigg]\,.
    \label{eq:Asimov_loglikelihood_test}
\end{align}
Fixing the value of the scalar mass, we then use a root-solving algorithm to find the value of $|g_{ee}|$ for which the test statistic $q_{\param^\phi} = 2.71$, corresponding to exclusion upper bound at a $90\%$ CL.

In what follows, we set the SM parameters $\param^{2\nu}$ to their central values given in \cref{tab:params}, since, as mentioned above, the uncertainties in $g_A^\text{eff}$ and $M_{GT}^{(-1)}$ cancel out. The effect of minimisation with respect to $\xi_{i1}$ and $\Delta_{0,2}$ on the resulting limits is subleading -- the largest uncertainty in that case comes from $\xi_{31}$~\cite{Morabit:2024sms}, the effect of which is negligible compared to the NME uncertainties (see \cref{fig:xi31unc}). We thus avoid minimisation to make the numerical computation faster. 

\subsection{Multi-observable analysis}
In the previous section, we considered a single observable $T$. However, current and next-generation experiments can measure multiple observables simultaneously, which may improve sensitivity to BSM signals. For instance, the full kinematic information of the two emitted electrons allows us to construct a multi-dimensional distribution in $\epsilon$ and $\cos\theta$, or even in individual electron energies $E_1$ and $E_2$, and $\cos\theta$,
 \begin{align}
     \frac{d\Gamma^{2\nu,\phi}}{d\epsilon \,d(\cos\theta)} &= \frac12\left(\frac{d\Gamma^{2\nu,\phi}}{d\epsilon} - \frac{d\Lambda^{2\nu,\phi}}{d\epsilon}\cos\theta\right)\,,\\
     \frac{d\Gamma^{2\nu,\phi}}{dE_1 dE_2 d(\cos\theta)} &= \frac12\left(\frac{d\Gamma^{2\nu,\phi}}{dE_1 dE_2} - \frac{d\Lambda^{2\nu,\phi}}{dE_1 dE_2}\cos\theta\right)\,.
 \end{align}

In general, for $n$ observables $T^{(j)}$, the expected number of events per $i$-th $n$-dimensional bin is given, analogously to~\cref{eq:1D-dist}, by
\begin{align}
    N_\text{exp}^{(i)} = \frac{N_\text{tot}}{\Gamma^{2\nu} + \Gamma^\phi}\int_{T_i^{(1)}}^{T_{i+1}^{(1)}} dT^{(1)}\dots \int_{T_i^{(n)}}^{T_{i+1}^{(n)}} dT^{(n)}\,\left(\frac{d^n\Gamma^\phi}{dT^{(1)}\dots dT^{(n)}}+\frac{d^n\Gamma^{2\nu}}{dT^{(1)}\dots dT^{(n)}}\right)\,.
\end{align}
The construction of the likelihood function then proceeds in the same way as in the single-observable analysis, but now the sum in~\cref{eq:Asimov_loglikelihood_test} is performed over multi-dimensional bins, which leads to more demanding numerical computations.

The improvement in sensitivity to $|g_{ee}|$ then depends on the degree to which the multi-dimensional BSM spectrum is distinct from the SM background compared to the single-observable case.

\section{\texorpdfstring{Sensitivity to the scalar-neutrino couplings}{Sensitivity to the scalar-neutrino couplings}}
\label{sec:results}

\begin{figure}[t!]
     \centering
     \begin{minipage}{.48\textwidth}
     \centering
     \includegraphics[width=\linewidth]{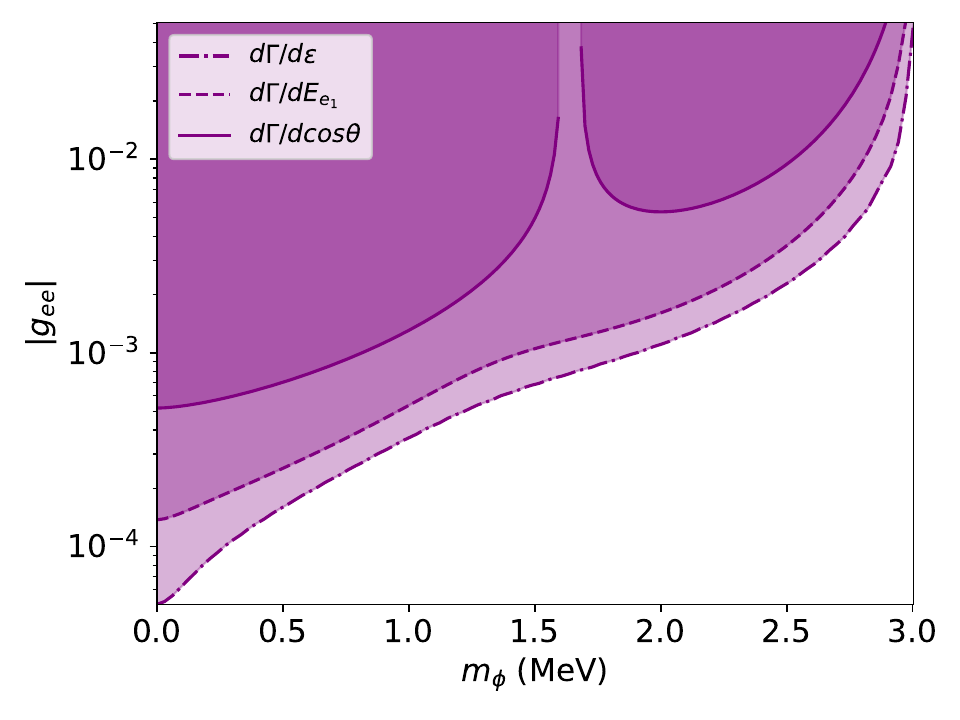}
     \caption{Sensitivity to the scalar-active neutrino coupling as a function of the scalar mass, for different observables for $^{100}$Mo.}
     \label{fig:angcorr}
     \end{minipage}
     \hfill
    \begin{minipage}{.48\textwidth}
     \centering
     \includegraphics[width=\linewidth]{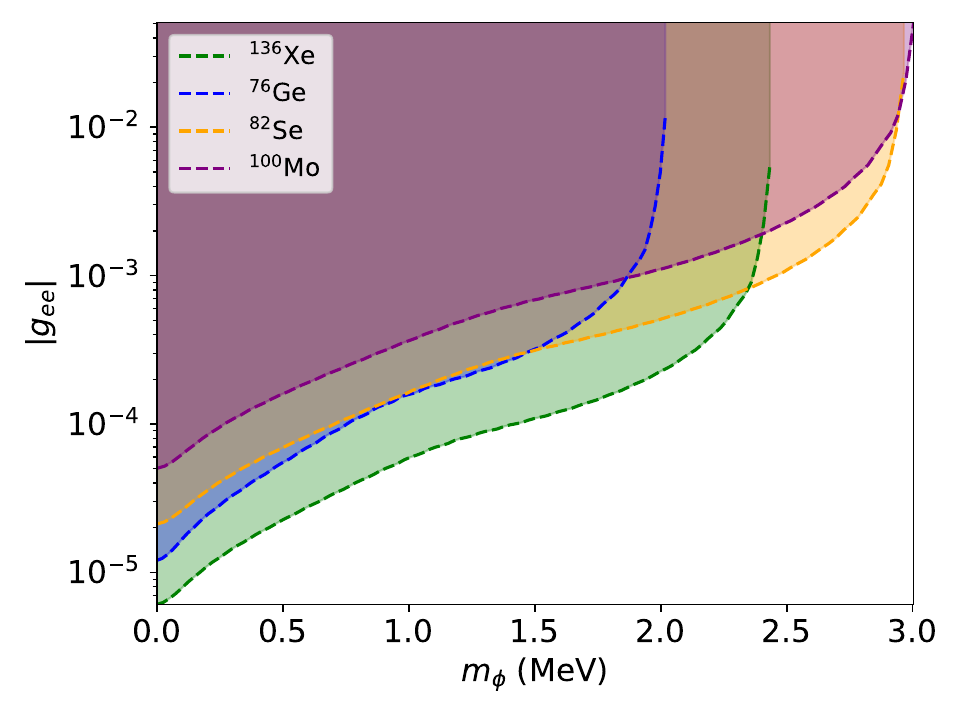}
     \caption{Sensitivity to the scalar-active neutrino coupling from the summed electron energy spectrum, for different isotopes.}
     \label{fig:isotopes}
     \end{minipage}
 \end{figure}

Following the statistical procedure outlined in the previous section, we can now compare the predicted BSM spectra against the SM spectrum, and put constraints on the BSM coupling of scalars to neutrinos for a given number of observed background events, fixed energy resolution, and assuming no BSM signal is seen. The primary observables are the summed electron energy, individual electron energies, and the angular correlation between them.  In \cref{fig:angcorr} we show the theoretical limits obtained for the three observables, for the isotope $^{100}$Mo, with $10^5$ events, an energy resolution of 100 keV, and an angular resolution $\Delta \cos\theta = 0.1$, inspired by the resolution and event count of latest experiments~\cite[e.g.,][]{NEMO-3:2019gwo}, and considering only SM \tvbb\;process as the background, ignoring systematic uncertainties. We note that increasing the resolution further has a miniscule effect on the limits obtained when compared to the effect of NME uncertainties (see \cref{fig:active_limit}). The sensitivity from the summed electron energy is the strongest in this case, and this applies to other isotopes as well for the chosen values of parameters. Hence, we will mainly show the results from the summed energy distribution, as well as use the event count and resolutions mentioned above, in the following. The angular observable fails to constrain the coupling for certain scalar masses due to $K^\phi$ being similar to that of \tvbb, as mentioned in \cref{sec:angular}.

\Cref{fig:isotopes} shows the limits from different isotopes under consideration; see \cref{app:ovbbamp} for the values of parameters used. We find that the higher order corrections to \tvbb, i.e., from the $G^{2\nu}_{i\neq0}$ terms and deviation of $\Delta_0$ from 1 in \cref{eq:2nugamma}, can account for up to a~relative $\sim20\%$ change in the limits on $|g_{ee}|$. Below its Q-value, \Xe\;provides the most stringent limit, and we will thus always show limits from \Xe\;unless explicitly mentioned otherwise. Beyond the Q-value of the isotopes, it is possible to obtain limits on the coupling by studying the spectral distortion caused in the \tvbb\;spectrum, caused now by a virtual scalar with $m_\phi>Q$ decaying into two neutrinos. This scenario is discussed further in \cref{sec:offshell}, and here we only consider scalars with mass below the Q-value of \Xe. 

\subsection{\texorpdfstring{Sensitivity to}{Sensitivity to} scalar coupling to active neutrinos} 
\begin{figure}[h!]
     \centering
     \begin{minipage}{.48\textwidth}
     \centering
     \includegraphics[width=\linewidth]{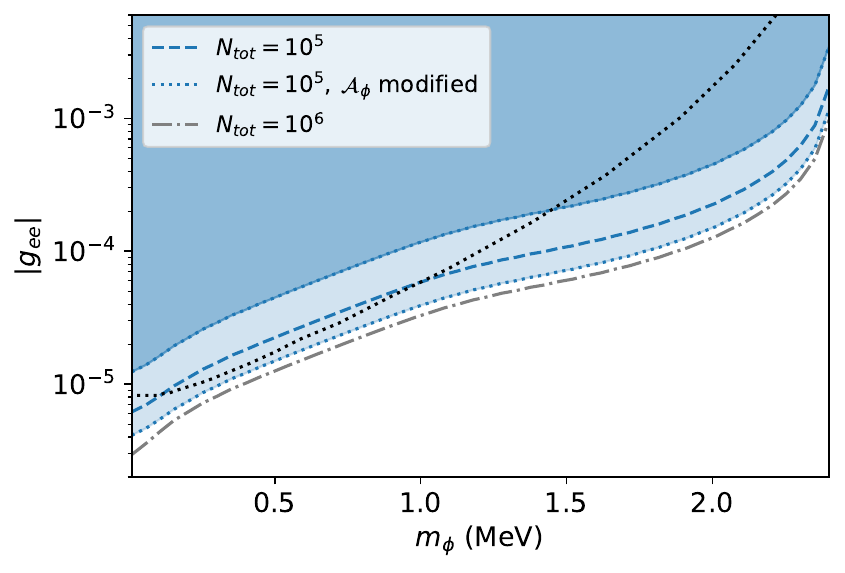}
     \caption{Variation in the sensitivity due to uncertainty in NMEs, shown as the light blue band; see text for details. The black line gives the limit from ref.~\cite{Brune:2018sab}.}
     \label{fig:active_limit}
     \end{minipage}
     \hfill
    \begin{minipage}{.48\textwidth}
     \centering
     \includegraphics[width=\linewidth]{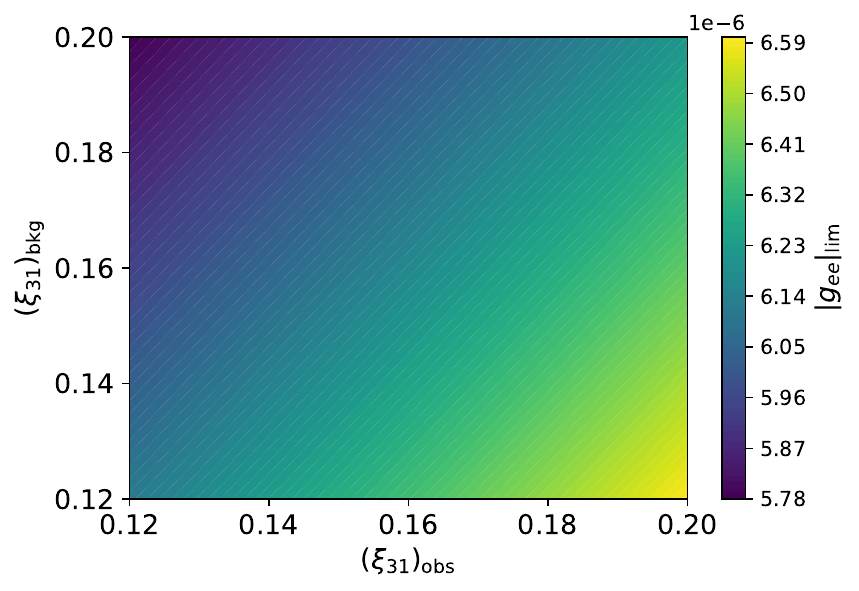}
     \caption{Change in the sensitivity to scalar-neutrino coupling when $\xi_{31}$ is allowed to vary independently for the background and the observed spectra.}
     \label{fig:xi31unc}
     \end{minipage}
 \end{figure}

We now turn to the first case discussed in \cref{sec:theory}, and as per the discussion in \cref{sec:stat}, put best possible theoretical limits, i.e., ignoring systematic and statistical uncertainties, and considering \tvbb\;decay as the only background, on the coupling $g_{ee} = \sum_{i,j}g_{ij}U_{ei}U_{ej}$. The limit, as a function of the scalar mass $m_\phi$, is shown in \cref{fig:active_limit}. The dashed line gives the limit for the central values of NMEs, and the light blue bands represent the uncertainties due to the overall NME in the \ovbbp\;amplitude when we change $\mathcal{A}^\phi(0)$ between $50\%$ and $100\%$ of its central value, with the edges shown with dotted lines. We set the number of events to $10^5$, and the energy resolution to $\sim 100$~keV. The grey dash-dotted line shows the improvement in limit if $10^6$ events are available for analysis. For comparison, the most optimistic limit from ref.~\cite{Brune:2018sab} is also given with the black dotted line.\footnote{Recently released experimental limits on massive scalar couplings from the PandaX collaboration can be found in ref.~\cite{PandaX:2025tls}.} Refs.~\cite{Brune:2018sab,Blum:2018ljv} indicate the systematic uncertainties based on the experimental limit for a massless scalar. A spectral analysis, as done here, would require more detailed detector simulations to obtain uncertainty estimates, rather than a simple recasting. Although the size of such uncertainty bands are expected to be similar for the present analysis, we do not include these uncertainties, but choose to focus on theoretical uncertainties, e.g., those due to NMEs, which turn out to be equally important.

For the low mass region, the small differences in the limit can be attributed to the fact that we have not included systematic uncertainties in our analysis (in particular for the massless case where the limit can be compared to ref.~\cite{EXO-200:2014vam}), in addition to the use of updated \ovbb\;computations. For larger scalar masses, however, the limit from ref.~\cite{Brune:2018sab} drops rather sharply in comparison, and the shape is also different. The analysis followed in refs.~\cite{Blum:2018ljv,Brune:2018sab} approximates the limit on massive scalars by rescaling the limit on massless scalars, based on the ratio of their phase space factors, as well as the maxima of the signal to background ratio, which can result in loss of information in the form of features in the spectral shapes. This explains the sharper drop in the limits compared to our results for larger scalar masses, where there is a relatively large phase space suppression. 

Apart from the uncertainties in the \ovbbp\;amplitude, there are uncertainties arising from the parameters in the background \tvbb\;process. We quantify these by varying $\xi_{31}$ between 0.12 and 0.20 in \emph{observed} and \emph{expected} spectra independently, and show the effect on the limit in \cref{fig:xi31unc}. We see that the change in the limit is much smaller here compared to the uncertainties due to the \ovbbp\;NMEs. Thus, in the following, we will set the value of the \tvbb\;parameters to the same value in the \emph{observed} and \emph{expected} spectra, i.e., make use of a representative Asimov dataset. 

 \paragraph{Using multiple observables.}

 With the upcoming experiments being able to resolve individual electron energies and angular correlations, it can be useful to look at multi-dimensional phase space distributions for deviations caused by BSM contributions; see \cref{sec:stat} for details. We find that a combined analysis for the summed electron energy $\epsilon$ and angular correlation $\cos\theta$ observables provides no improvement to the limits. This can be explained by the fact that the electron angular correlation for the scalar emitting decay is rather similar to \tvbb. However, as shown in \cref{sec:rhc}, a combined analysis for the case of right-handed currents leads to an improvement in limits compared to an analysis for just the summed energy.
 
\subsection{\texorpdfstring{Sensitivity to}{Sensitivity to} scalar coupling involving sterile neutrinos} 
\begin{figure}[h!]
     \centering
     \begin{minipage}[t]{0.48\textwidth}
     \centering  \includegraphics[width=\linewidth]{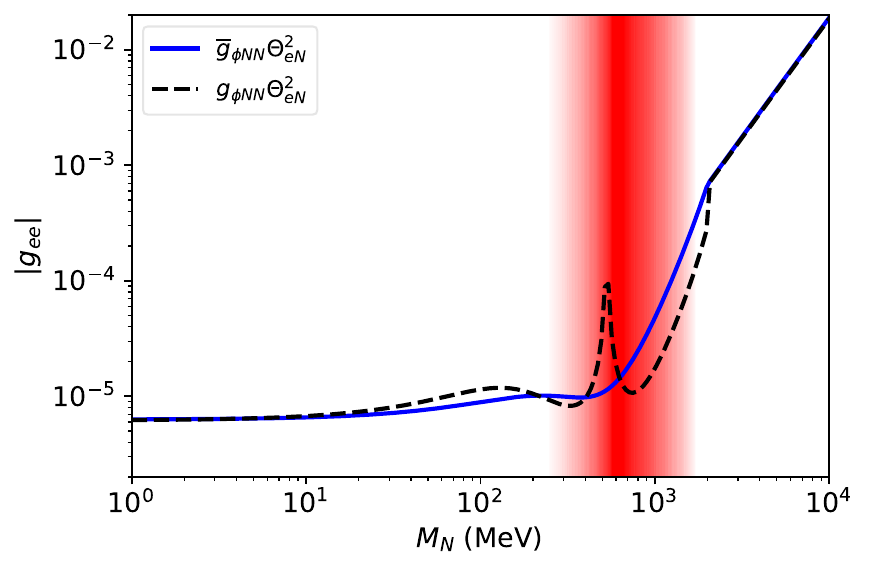}
     \caption{Sensitivity to a massless scalar coupling to sterile neutrinos, as a function of sterile mass. The scalar coupling exhibits a funnel, and the position of the funnel can lie within the region shaded in red.}
     \label{fig:sterile_limit}
     \end{minipage}
     \hfill
     \begin{minipage}[t]{0.48\textwidth}
     \centering
     \includegraphics[width=\linewidth]{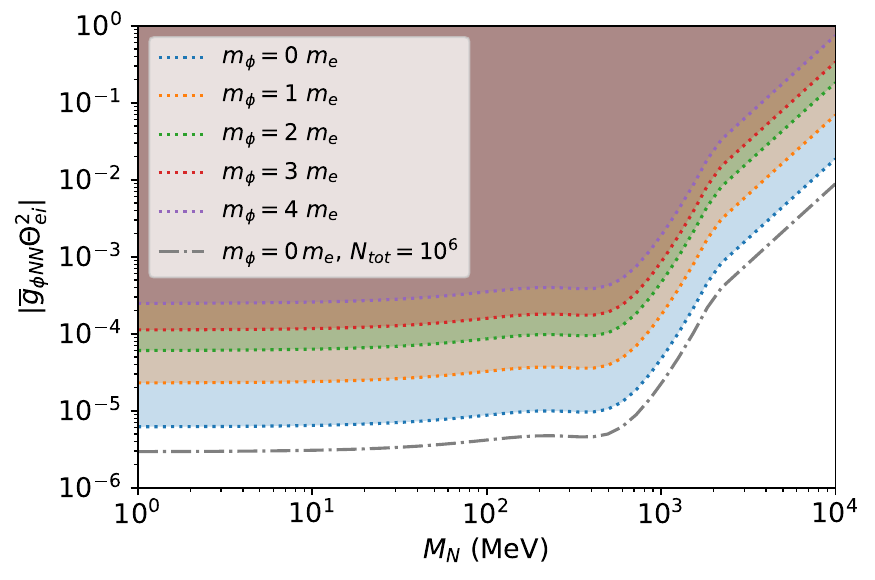}
     \caption{Sensitivity to the pseudoscalar coupling to sterile neutrinos for different scalar masses. The grey dash-dotted line shows the improvement in limits with $10\times$ more data points.}
     \label{fig:sterile_massive_pseudoscalar}
     \end{minipage}
 \end{figure}

In the scenario where the scalar couples only to sterile neutrinos, we consider the couplings \stc\;and 
\stcg\;one at a time, and show the limits in \cref{fig:sterile_limit} as a function of the sterile neutrino mass, for a massless scalar. While the behaviour of the limit on \stcg\;is fairly simple, the limit on \stc\;exhibits a funnel. This can be attributed to the sign flip in the amplitude going from the low mass region to the high mass region, beyond 2~GeV. The amplitude then necessarily needs to pass through zero in order to remain continuous. For the mass corresponding to the crossing, the rate will vanish, and we cannot put constraints on the coupling. The exact location of this funnel is dependent on the fit parameters used as it can lie almost anywhere within the region shaded in red, and we caution that the position shown in \cref{fig:sterile_limit} should be considered as only a representative picture. The envelope of the limit, excluding the funnel, should however remain fairly robust.

\Cref{fig:sterile_massive_pseudoscalar} shows the limit on the coupling \stcg, but now for different masses of the scalar. The limit on \stc\;will be similar but with a region between 300~MeV and 1~GeV where a funnel would be present, as can be expected from \cref{fig:sterile_limit}. Once again, the limits become weaker for larger scalar masses as the peak of the spectrum moves to lower energies, while also getting weaker, with increasing $m_\phi$. In grey (dash-dotted), the improvement in limits with an order of magnitude improvement in the number of available data points is shown.

\paragraph{Mixed couplings.} The NMEs in this scenario are identical to the case of the pseudoscalar couplings of scalar-sterile neutrino interactions. As a result, the limit on the effective coupling $g_{ee}=\Theta_{eN}\sum_i U_{ei}g_{\phi N \nu_i}$, as discussed in \cref{sec:theory}, can be read off from \cref{fig:sterile_massive_pseudoscalar}. 

\section{Off-shell scalar decaying into neutrinos}
\label{sec:offshell}

\begin{figure}[t!]
    \centering
    \begin{tikzpicture}[scale=0.9, transform shape]
\begin{feynman}
\vertex (a); 
\node[blob,fill=black] (bl) at (a) ;
\vertex [above right=of a] (b);
\vertex [right=of b] (c);
\vertex [right=of c] (d);
\vertex [above right=1.2 of d] (e1);
\vertex [below right=0.8 of d] (nu1);
\vertex [right=of d] (e);
\vertex [below right=of e] (f);
\node[blob,fill=black] (bl) at (f);
\vertex [below right=of a] (g) ;
\vertex [right=of g] (v2);
\vertex [right=of v2] (t2);
\vertex [below right=1.2 of t2] (e2);
\vertex [above right=0.8 of t2] (nu2);
\vertex [below left=of f] (h) ;
\vertex [above=1 of v2] (m1) ;
\vertex [right=1 of m1] (m2) ;
\diagram* {
(a) -- [quarter left, edge label = \(n\)] (b) -- [fermion] (e) -- [quarter left, edge label = \(p\)] (f),
(a) -- [quarter right, edge label'= \(n\)] (g) -- [fermion] (h) -- [quarter right, edge label' = \(p\)] (f),
(v2) -- [fermion] (e2),
(c) -- [fermion] (e1),
(v2) --  (c),
(m1) -- [scalar, edge label' = \(\phi\)] (m2),
(m2) -- (nu1),
(m2) -- (nu2),
};
\vertex[right= 0.1 of e2]{\(e^-\)};
\vertex[right= 0.1 of e1]{\(e^-\)};
\vertex[right= 0.1 of nu2]{\(\nu_k\)};
\vertex[right= 0.1 of nu1]{\(\nu_l\)};
\vertex[blob,fill=gray,scale=0.5] at (c) {};
\vertex[blob,fill=gray,scale=0.5] at (v2) {};
\end{feynman}
\end{tikzpicture}
    \caption{A heavy off-shell scalar decaying into two neutrinos.}
    \label{fig:off_shell_feynman}
\end{figure}
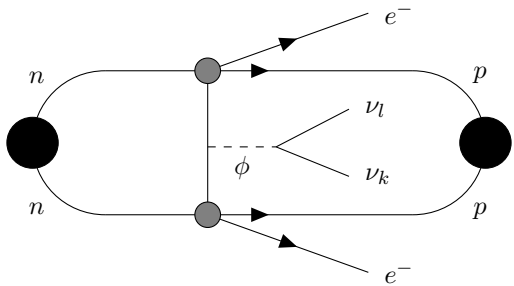

Although double beta decay primarily constrains scalars that can be produced on-shell in the final state, one can also consider the case of a scalar with mass greater than the Q-value of the isotope under consideration, such that it is produced off-shell and decays into two neutrinos as shown in \cref{fig:off_shell_feynman}, thus distorting the original \tvbb\;spectrum~\cite{Deppisch:2020sqh,Berryman:2022hds}.

For simplicity, let us first consider the case with active neutrinos, and consider only $g_{ij}$; a replacement $g_{ij} \rightarrow g_{ij}+\bar g_{ij}$ in the following will let us recover the complete scenario. The $s$-channel diagram, where the scalar momentum is negligible compared to the exchange neutrino momentum, for this decay will modify the on-shell scalar emission diagram with the spinor line for the external neutrinos coupled to the scalar, the coupling $\sum_{k,l}g_{kl}$, and the scalar propagator. In the scalar propagator, the momentum carried by the scalar will equal the sum of external momenta of the neutrinos. Focusing on scalar masses well above the $Q$ value, the resulting phase space factor will modify the \tvbb\;rate as
\begin{align}
    \Gamma_\text{tot}\propto &\int dE_{e_1} dE_{e_2} dE_{\nu_1} dE_{\nu_2} E_{e_1} E_{e_2} |\vect p_{e_1}||\vect p_{e_2}| E_{\nu_1}^2E_{\nu_2}^2 \delta\left(Q + 2m_e - E_\text{K}\right) F(Z_f, E_{e_1})F(Z_f,E_{e_2})\nonumber\\
    &\times \Bigg[1 + \mathcal{A}_2 \xi_{31}\frac{\Delta_2}{\Delta_0} + \mathcal{A}_4\left(\xi_{51}\frac{\Delta_2}{\Delta_0} + \frac13\xi_{31}^2\right)+\mathcal{A}_{22}\frac13\xi_{31}^2+\frac{\mathcal{A}_M}{\Delta_0}\nonumber\\
    &+\left|\frac{\mathcal{A}^\phi}{M_{GT}^{(-1)}}\right|^2\frac{1}{\Delta_0}\left(\frac{m_e}{4\pi R}\right)^2\frac{1}{m_\phi^4 - m_\phi^2\Gamma_\phi^2}\left|\sum_{i,j}g_{ij}U_{ei}U_{ej}\sum_{k,l}g_{kl}\right|^2 \nonumber\\
    &- 2\, \frac{\mathcal{A}^\phi}{M_{GT}^{(-1)}}\frac{1}{\Delta_0^{\frac12}}\,\frac{m_e}{4\pi R}\,\text{Re}\left(\frac{1}{- m_\phi^2 + i\,m_\phi\Gamma_\phi}\sum_{i,j}g_{ij}U_{ei}U_{ej}\sum_{k,l}g_{kl}\right)\Bigg]\,,
    \label{eq:offshelltonunu}
\end{align}
where $E_\text{K}$ is the sum of kinetic energies of the emitted electrons and neutrinos. In the above formula, we have employed the results of ref.~\cite{Deppisch:2020sqh}, but, as before, we also included here the state-of-the-art corrections to \tvbb\;rate~\cite{Simkovic:2018rdz,Morabit:2024sms}. More specifically, the decay width for $\phi\rightarrow \nu\nu$ is in this case given by $\Gamma_\phi = \frac{m_\phi}{8\pi}|\sum_{i,j} g_{ij}|^2$, where an additional symmetry factor $\frac12$ appears for $i=j$, and the masses of the neutrinos are considered to be negligible.  The second line contains the leading order term and the corrections to \tvbb\;rate discussed in ref.~\cite{Morabit:2024sms}, the third line contains the contribution to the rate from the scalar decaying to two neutrinos, and the fourth line is the interference between the two pieces. The constant of proportionality can be read off from \cref{eq:2nugamma,eq:2nupsf}. The relative factors containing $m_e/(\pi\,R)$ arise due to the definitions of the normalised amplitudes; cf. \cref{eq:ampfactor}. 
The factor 
\begin{align}\left|\frac{\mathcal A_\phi}{M_{GT}^{(-1)}}\frac{1}{\Delta_0^{1/2}}\frac{m_e}{4\pi R}\right|^{1/2} \frac{1}{m_\phi}\simeq \frac{20\,\mathrm{MeV}}{m_\phi}
\end{align} (where the numerical evaluation is taken for \Xe) multiplied by the scalar-neutrino coupling strength determines the relative size of the new physics contribution compared to the \tvbb\;rate.

\begin{figure}[t!]
    \centering
    \includegraphics[width=0.6\linewidth]{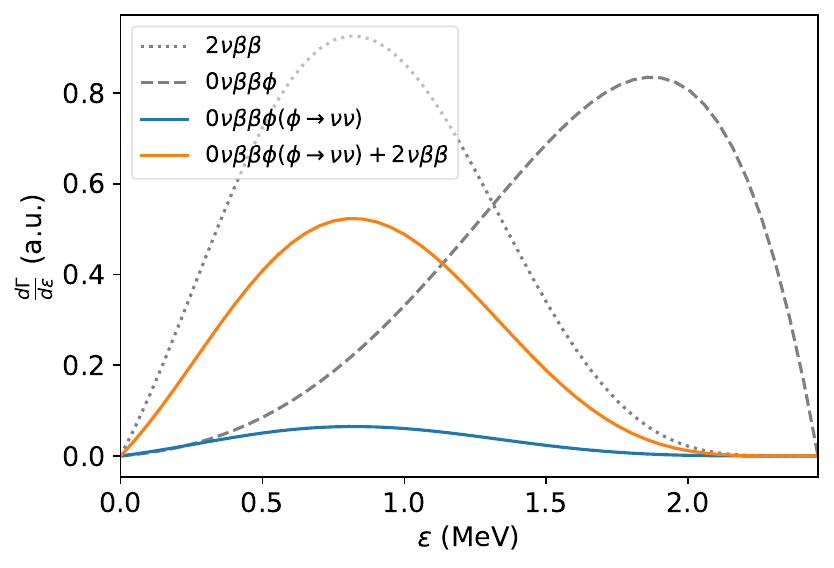}
    \caption{Size of the phase space distributions for heavy scalars decaying into active neutrinos from \cref{eq:offshelltonunu}, for $m_\phi = 5$~MeV, $\sum_{i,j}g_{i,j} = \sum_{i,j}g_{i,j}U_{ei}U_{ej}=0.1$, and $g_{ij}=0$ if $i\neq j$. The blue and orange lines are from the scalar decay and the interference term, respectively. They are normalised to the \tvbb\;rate, and magnified by $10\times$ for visualisation. Normalised \tvbb\;and massless on-shell scalar emitting spectra are shown for reference.}
    \label{fig:offshell_spectra}
\end{figure}

\Cref{fig:offshell_spectra} gives the summed energy distributions for an off-shell scalar decaying to two active neutrinos for $m_\phi = 5$~MeV. The regular \tvbb\;and \ovbbp\;spectra are shown with grey lines, and are normalised to unity. We have picked $\sum_{i,j}g_{i,j} = \sum_{i,j}g_{i,j}U_{ei}U_{ej}=0.1$, and $g_{ij}=0$ if $i\neq j$, for the purpose of visualisation, and the sign of the NMEs are kept as mentioned in \cref{tab:params,app:ovbbamp}. The scalar-induced terms are magnified by $10\times$ to enhance the visibility. The shape of the spectra are considerably different from the usual scalar emitting case.

This phase space distribution can be compared against the \tvbb\;background, i.e., the $g_{ij}\rightarrow0$ limit, to obtain limits on the effective coupling $|\sum_{i,j}g_{ij}U_{ei}U_{ej}||\sum_{k,l}g_{kl}|$ in this case. However, on top of the effective coupling being quadratic in the scalar-neutrino coupling $g_{ab}$, the limit is further weakened by the phase space suppression and the heavy scalar propagator.
A $t$-channel diagram could be also considered; however, it would be suppressed compared to $s$-channel decay due to an additional propagator, and we do not expect the contribution to significantly modify the limits obtained here.

We now consider the case of a massive off-shell scalar decaying into light sterile neutrinos. If the light sterile neutrino mass is sufficiently below the Q-value, there are several possible sources of distortion to the \tvbb\;spectrum. First of all, the decay may involve the emission of one active and one sterile neutrino via mixing at the two beta decay vertices. Secondly, both vertices may involve emission of sterile neutrinos. Finally, this diagram can interfere with the \ovbbp\;diagram where the scalar then mediates a sterile neutrino self-interaction resulting in a four-lepton final state. Once again, we ignore the $t$-channel diagram contribution to the two-sterile final state. The total rate then reads
\begin{align}
    \Gamma_\text{tot} = \,&\left(1-|\Theta_{eN}|^2\right)^2\,\Gamma^{2\nu} + 2\left(1-|\Theta_{eN}|^2\right)|\Theta_{eN}|^2\,\Gamma^{\nu N} + |\Theta_{eN}|^4\,\Gamma^{2N}\,,
    \label{eq:lightsterilerate}
\end{align}
where $\Gamma^{2\nu}$ is the SM \tvbb\;rate given in \cref{eq:2nugamma}. The single sterile neutrino emitting rate $\Gamma^{\nu N}$ is given similarly in terms of the phase space integrals $G^{\nu N}_i$, but now without the magnetic piece,
\begin{align}
    G_i^{\nu N} = \,&\frac{1}{\ln 2}\frac{\left(G_F\,V_{ud}\right)^4}{8\pi^7\,m_e^2}\int_{m_e}^{Q+ m_e-m_N} dE_{e_1} \int_{m_e}^{Q+2m_e - E_{e_1} - m_N} dE_{e_2} \int_0^{Q+2m_e - E_{e_1} - E_{e_2}-m_N} dE_{\nu} \nonumber\\
    &\times F(Z_f, E_{e_1})F(Z_f, E_{e_2})E_{e_1}|\vect p_{e_1}|E_{e_2}|\vect p_{e_2}|E_{\nu}^2E_{N}\sqrt{E_N^2 - m_N^2}\,  \mathcal{A}_i^{2\nu}\, ,
    \label{eq:nuNPSF}
\end{align}
where $E_{N} = E_i - E_f - E_{e_1} - E_{e_2} - E_{\nu}$. We consider neither mixed couplings of the scalar here, nor the scalar-active neutrino coupling, and hence, scalar-mediated $\nu\nu/\nu N$ decays do not play a part in the rates.

The rate for the double sterile neutrino emitting process is more involved due to the interference between $2N\beta\beta$ and $\ovbbp(\phi\rightarrow NN)$ diagrams. We write it as
\begin{align}
    \Gamma^{2N} \propto \,&\int_{m_e}^{Q+ m_e-2m_N} dE_{e_1} \int_{m_e}^{Q+2m_e - E_{e_1}-2m_N} dE_{e_2} \int_{m_N}^{Q+2m_e - E_{e_1} - E_{e_2}-m_N} dE_{N_1} \nonumber\\
    &\times F(Z_f,E_{e_1})F(Z_f,E_{e_2})E_{e_1}|\vect p_{e_1}|E_{e_2}|\vect p_{e_2}| \nonumber\\
    &\times E_{N_1}\sqrt{E_{N_1}^2 - m_N^2}E_{N_2}\sqrt{E_{N_2}^2 - m_N^2}\,  \mathcal{A}^{2N}\,,
    \label{eq:NNPSF}
\end{align}
with $E_{N_2} = E_i - E_f - E_{e_1} - E_{e_2} - E_{N_1}$. $\mathcal{A}^{2N}$ can be read off from the square brackets in \cref{eq:offshelltonunu}, once again without the magnetic piece, with the replacement to the effective coupling factor $\sum_{i,j}g_{ij}U_{ei}U_{ej}\sum_{k,l}g_{kl}\rightarrow g_{\phi NN}^2$ (note that the mixing $\Theta_{eN}$ has already been factored out in \cref{eq:lightsterilerate}). For large $m_\phi$, at leading order, the sterile neutrino self-interaction term is just $\frac{g_{\phi NN}^2}{m_\phi^2}N^4$, and comparing the total decay rate against the SM expectation, one can constrain the strength of this higher-dimensional operator, or equivalently, put lower bounds on the scalar mass $m_\phi$ as a function of $m_N$, $g_{\phi NN}$, and $|\Theta_{eN}|$. Such an interaction has been discussed, for instance, in the context of self-interacting sterile neutrino dark matter~\cite{Johns:2019cwc,Fuller:2024noz}.

\section{Right-handed beta decay currents}
\label{sec:rhc}

Scalar emission in \ovbb\;may also proceed through higher-order effective vertices involving right-handed currents. We consider the case of a vertex with left-handed quark currents, and right-handed lepton currents. Following ref.~\cite{Cepedello:2018zvr}, we parametrise the vertex as
\begin{align}
    \mathcal L \supset \frac{G_F\,V_{ud}}{\sqrt{2}}\frac{\epsilon_{LR}}{m_N}\,J_\mu^L\,j^\mu_R\,\phi\,,
    \label{eq:rhc_vertex}
\end{align}
for the lepton and quark currents $j^\mu$ and $J^\mu$ respectively. The $m_N$ suppression keeps the coupling $\epsilon_{LR}$ dimensionless.

In the EFT prescription of ref.~\cite{Dekens:2020ttz}, this vertex can be mapped to the $C_{VLR}^{(6)}$ vertex. For \ovbbp, we consider one insertion of the regular beta decay coupled with one insertion of the scalar emitting vertex. This leads to an effective neutrino potential of the form shown in ref.~\cite[eq.~(93)]{Dekens:2020ttz}, with one insertion of $C_{VLL}^{(6)}$ and $C_{VLR}^{(6)}$ each. Note that in our case, however, the overall factor of $m_N$ cancels against the $m_N$ in the definition of the vertex, and the term linear in difference in electron energies is also neglected. We also choose to absorb the relative factor of $\frac12$ in the definitions of the vertices into the phase space factor given below.

The resulting decay proceeds as shown in \cref{fig:rhc_feynman}, and the decay rate is then given by 
\begin{align}
    \frac{\Gamma^\phi}{\ln 2}=\overline \Omega^\phi\left|\sum_i\left(\epsilon_{LR}\right)_{ei}U_{ei}\right|^2 |\bar{\mathcal A}^\phi|^2\,,
\end{align}
where the phase space factor is~\cite{Cirigliano:2017djv}
\begin{align}
    \overline \Omega^\phi \propto &\int dE_{e_1}\,dE_{e_2}\,dE_\phi \,d\cos\theta\,|\vect{p}_{e_1}||\vect{p}_{e_2}||\vect p_\phi|\,\delta (E_i-E_f-E_{\phi}-E_{e_1}-E_{e_2})\nonumber\\
    &\times F(Z_f,E_{e_1})F(Z_f,E_{e_2})\, \left(E_{e_2}E_{e_1}+m_e^2+\vect p_{e_1}\cdot \vect p_{e_2}\right)\,,
\end{align}
where the constant of proportionality is the same as in \cref{eq:majoronPSF}. The final, angle-dependent, term does not contribute to the total rate. The opposite sign of this term compared to \cref{eq:angular} makes an angular analysis particularly useful, given the opposite trend of the differential rate with $\cos \theta$.

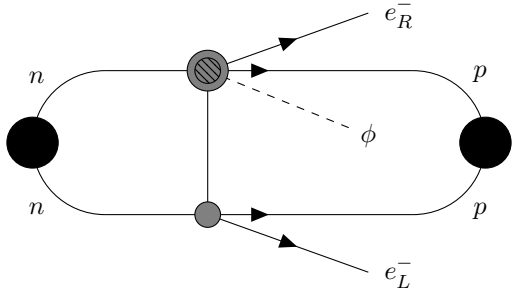
\begin{figure}
    \centering
    \begin{tikzpicture}
        \begin{feynman}[scale=0.9, transform shape]
\vertex (a); 
\node[blob,fill=black] (bl) at (a) ;
\vertex [above right=of a] (b);
\vertex [right=of b] (c);
\vertex [right=of c] (d);
\vertex [above right=1.2 of d] (e1);
\vertex [right=of d] (e);
\vertex [below right=of e] (f);
\node[blob,fill=black] (bl) at (f);
\vertex [below right=of a] (g) ;
\vertex [right=of g] (v2);
\vertex [right=of v2] (t2);
\vertex [below right=1.2 of t2] (e2);
\vertex [below left=of f] (h) ;
\vertex [below=1.5 of e1] (m1) {\(\phi\)};
\diagram* {
(a) -- [quarter left, edge label = \(n\)] (b) -- [fermion] (e) -- [quarter left, edge label = \(p\)] (f),
(a) -- [quarter right, edge label'= \(n\)] (g) -- [fermion] (h) -- [quarter right, edge label' = \(p\)] (f),
(v2) -- [fermion] (e2),
(c) -- [fermion] (e1),
(v2) --  (c),
(c) -- [scalar] (m1),
};
\vertex[right= 0.1 of e2]{\(e^-_L\)};
\vertex[right= 0.1 of e1]{\(e^-_R\)};
\vertex[blob,fill=gray,scale=0.8] at (c) {};
\vertex[blob,scale=0.5] at (c) {};
\vertex[blob,fill=gray,scale=0.5] at (v2) {};
\end{feynman}
    \end{tikzpicture}
    \caption{\ovbbp\;proceeding via the right-handed vertex given in \cref{eq:rhc_vertex}, which is shown with the bigger, hatched, grey circle. The subscript on the electrons denotes their handedness; unlike \cref{fig:ovbb_feynman}, where both of the emitted electrons are left-handed, the scalar-emitting vertex now gives a right-handed electron in the final state.}
    \label{fig:rhc_feynman}
\end{figure}

\begin{figure}[t!]
    \centering
    \begin{subfigure}[b]{0.49\textwidth}
        \includegraphics[width=\linewidth]{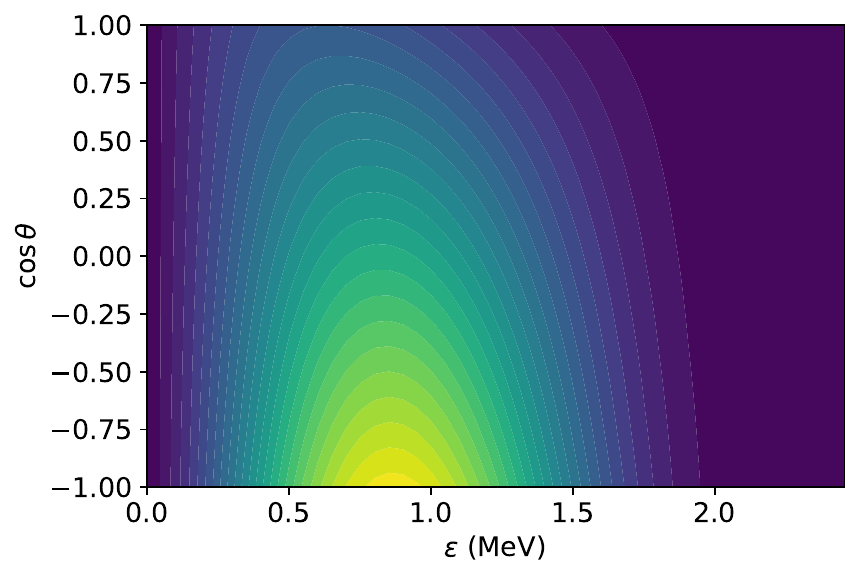}
    \end{subfigure}
    \begin{subfigure}[b]{0.49\textwidth}
        \includegraphics[width=\linewidth]{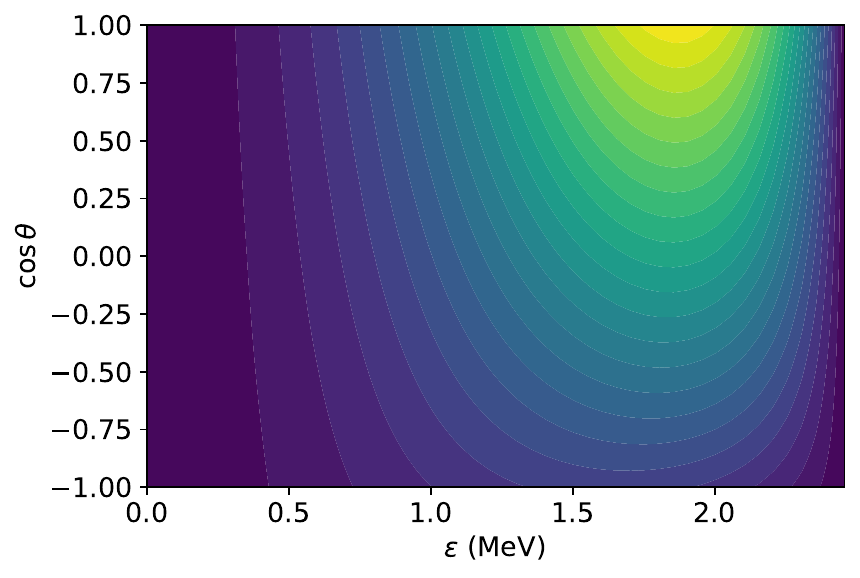}
    \end{subfigure}
    \caption{Normalised phase space distribution in the summed electron energy $\epsilon$ and angular correlation $\cos\theta$ for \tvbb\;decay (left) and the right-handed current induced scalar-emitting decay, for a massless scalar, discussed in \cref{sec:rhc} (right). Lighter colours denote higher relative strength of the spectrum.}
    \label{fig:2Dspectra}
\end{figure}

\Cref{fig:2Dspectra} shows the phase space distributions for the regular \tvbb\;decay and the (massless) scalar-emitting decay via a right-handed lepton current. The spectra are normalised to unity, and lighter colours denote higher strengths. The \tvbb\;spectrum prefers back-to-back emission of electrons, as shown in \cref{fig:spectra_angular}, and peaks at $\epsilon\sim 1$~MeV. The scalar-emitting spectrum in this case has a summed energy distribution similar to the massless scalar case in \cref{fig:spectra}, as was also shown in ref.~\cite{Cepedello:2018zvr}, but prefers collinear electron emission. Given this vastly different behaviour compared to the background \tvbb\;decay, an analysis of the angular correlation between the electrons could provide valuable information when these kinds of decays are involved.

The modified amplitude for this decay,\footnote{This NME is comparable with the function $a\left(E_1,E_2\right)$ in ref.~\cite{Cepedello:2018zvr}. We, however, ignore the higher-order, energy-dependent terms here.} in the massless neutrino limit, is~\cite{Dekens:2020ttz}
\begin{align}
    \bar{\mathcal A}^\phi = &-\frac{g_A}{g_M}\left[\mathcal M_{GT}^{MM} + \mathcal M_T^{MM}\right] + \frac{m_\pi^2}{m_N\Lambda_\chi}\left[\frac{1}{g_A^2}g_{V}^{NN}\mathcal M_{F,sd}-\frac14 g_{V}^{\pi N}\left(\mathcal M^{AP}_{GT,sd} + \mathcal M^{AP}_{T,sd}\right)\right]\,,
\end{align}
where the first term corresponds to the potential contributions, while the second term is the short-distance contribution. The LECs $g_{V}^{\pi N,NN}\sim \order{1}$ by naive dimensional analysis, and we set them to 1. For the ``magnetic'' NMEs, we use $\mathcal M_{GT}^{MM}=0.19$ and $\mathcal M_{T}^{MM}=0$~\cite{Menendez:2017fdf}, and for the short-distance NMEs, we use the values given in \cref{app:ovbbamp}. Only magnetic NMEs enter the potential contribution in this case as the leading order term in the currents vanishes, and only the higher order spin-dependent terms survive.

\begin{figure}[t!]
    \centering
    \includegraphics[width=0.7\linewidth]{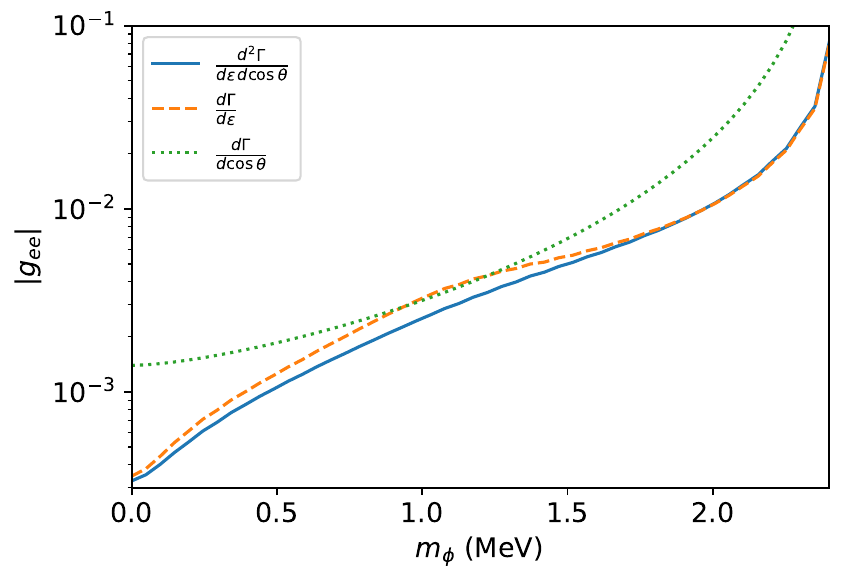}
    \caption{Sensitivity to the effective coupling involving one right-handed beta decay current. The limits from multi-observable analysis and single observable analysis are shown in solid and dotted (dashed) lines.}
    \label{fig:rhclim}
\end{figure}

The decay amplitude is much smaller compared to the standard case with left-handed currents, leading to relatively weaker limits on the effective coupling $g_{ee} = \sum_i \left(\epsilon_{LR}\right)_{ei}U_{ei}$, as shown in \cref{fig:rhclim}. The solid line shows the limit drawn from a multi-observable analysis with the summed electron energy and angular correlation. The dashed and dotted lines show the limits obtained when only one of the observables is analysed. We see that a study of the angular correlation leads to an improvement of the limits around $m_\phi \sim1$~MeV in this case, owing to the opposite sign compared to \cref{eq:angular}, although the improvement is marginal and falls within the range of uncertainties. For low and high scalar masses, the angular correlation does not provide much extra information, and thus the limit is essentially set by the difference in the summed electron energy spectra.

Similar analyses can be performed for various different types of exotic currents, which may benefit from a study of a combination of variables rather than a single observable analysis of the summed electron energy. For example, if the hadronic current in \cref{eq:rhc_vertex} is also right-handed, the amplitude vanishes at leading order unless the electron p-waves are also considered~\cite{doi:1985dx,Cepedello:2018zvr}, and the higher order term that is proportional to the difference in the electron energies remains, i.e., $\bar{\mathcal{A}}^\phi\propto (E_{e_1} - E_{e_2})$ in this case. This gives rise to a~distinct behaviour in the spectrum in the $E_{e_1}-E_{e_2}$ plane (see ref.~\cite{Cepedello:2018zvr}), which can help distinguish the signal from the background better than a single observable study focussing on the total kinetic energy of the electrons. However, since the amplitude is suppressed in such scenarios, the overall limit obtained will be weaker than the standard case of scalar emission. At the same time, the improvement is substantial only for decay with emission of a scalar with non-zero masses, for which the summed electron energy spectrum peaks around the same energy as for standard \tvbb\;decay, making the associated constraint from the summed energy spectra weaker.

\section{Conclusions}
\label{sec:conclusion}

Neutrinoless double beta decay would be a definitive indicator of new physics and the Majorana nature of neutrinos, characterized by a relatively clean signal at the Q-value of the decaying isotope. However, if the decay is accompanied by the emission of a scalar particle, the spectrum becomes distorted, making the signal harder to distinguish from the background of two-neutrino double beta decay. Often studied in the context of Majorons, neutrinoless double beta decay with the emission of additional scalars opens new avenues for probing beyond Standard Model (BSM) physics. In this work, we revisited this process in light of recent improvements in the predictions of neutrinoless double beta decay rates, also considering the possibility of the emitted scalar being massive particles.

We used the state-of-the-art theoretical description of neutrinoless and neutrinofull double beta decay to improve the description of scalar emission in double-weak decays in various scenarios. We focused on the theoretical description and did not consider possible experimental uncertainties and backgrounds. As shown in \cref{fig:angcorr,fig:isotopes}, the optimal observable for obtaining the best limits to the coupling of such scalars to neutrinos is the summed electron energy, and \Xe\;provides the strongest limits, up to its Q-value, among the isotopes we considered. We found that while the uncertainties in higher order matrix elements in \tvbb, parametrized with the ratios $\xi_{i1}$, affect the obtained limits, the major source of uncertainty comes from the \ovbb\;decay amplitude which contains several LECs and NMEs that are yet to be determined precisely. However, in essence these rescale the limits of the BSM scalar couplings and does not change the qualitative picture.

We extended the study to the case of scalar coupled to sterile neutrinos, with more general masses. Depending on the nature of the coupling, scalar or pseudoscalar, the amplitude of this decay may deviate from the \ovbb\;case. It is seen that for scalar couplings, there exists a mass range of sterile neutrinos where the limit obtained drops rapidly, introducing a ``funnel'', as the amplitude undergoes a sign flip. The limit saturates to the same value as for active neutrinos for sterile neutrino masses $\lesssim$~MeV, above which it starts getting weaker.

Scalars heavier than the Q-value of the isotope were also studied through their decay into two on-shell neutrinos. The resulting limits are expected to be relatively weaker for scalar masses well beyond the Q-value, as expected due to the extra scalar propagator, as well as the insertion of two scalar-neutrino vertices in the decay. Finally, an exotic scalar-emitting ``right-handed'' effective vertex was studied to demonstrate how a multi-observable analysis can lead to improvement in the limits to the effective coupling. The improvement is found to be within the range of NME uncertainties; therefore, more work on that front would be required to obtain more precise limits.

\subsection*{Note}

During the finalisation of this manuscript, a similar analysis~\cite{Boudjema:2025okq}, and new experimental limits~\cite{PandaX:2025tls} appeared online. 

\acknowledgments

We thank Wouter Dekens and Saad el Morabit for useful discussions and clarifications, and Frank F. Deppisch for comments on the final manuscript. JdV
acknowledges support from the Dutch Research Council (NWO) in the form of a VIDI grant. LG acknowledges
support from the Dutch Research Council (NWO) under
project number VI.Veni.222.318 and from Charles University through project PRIMUS/24/SCI/013. VP thanks the Mainz Institute for Theoretical Physics (MITP) of the Cluster of Excellence PRISMA+ (Project ID 390831469) and IPNP at Charles University, Prague for their hospitality during part of this project. DS acknowledges support from the Charles University through grants PRIMUS/24/SCI/013 and UNCE/24/SCI/016 and through the Charles University Mobility Fund. He is also grateful to Nikhef for hospitality provided during the initial phase of this project.

\appendix

\section{\texorpdfstring{\ovbb\;amplitudes}{Neutrinoless double beta decay amplitudes}}
\label{app:ovbbamp}

The NMEs of the \ovbbp\;process are closely related to \ovbb\;NMEs, and we briefly outline here the computation of these amplitudes as discussed in refs.~\cite{Dekens:2023iyc,Dekens:2024hlz,Cirigliano:2024ccq}. The mass-dependent amplitude is given by
\begin{eqnarray}\label{eq:fullint}
\amp(m_i) = \begin{cases}
\amp^{\rm (ld)}(m_i)+\amp^{\text{(sd)}}(m_i)\,,& m_i <  2 \text{ GeV}\,,\\
\amp^{\text{(9)}}(m_i)\,,& {\rm }\,  2 \text{ GeV} \le m_i \,,
\end{cases}
\end{eqnarray}
where $\amp^\text{(sd,ld)}$ stand for the short- and long-distance contributions. Note that here we ignore the ultrasoft contributions, which provide subleading corrections since, in the cases we study, the cancellation due to $\sum_i U_{ei}^2m_i = 0$ does not hold.

For large neutrino masses, the neutrino is integrated out at the quark level, and the amplitude stemming from the dimension-9 operator, with $\mu_0\simeq 2$ GeV, is given by
\begin{align}
\amp^{(9)}(m_i) = -2\eta(\mu_0,m_i) \frac{m_\pi^2}{m_i^2}\Bigg[&\frac{5}{6}g_1^{\pi\pi}\left(\mathcal{M}_{GT,sd}^{PP}+\mathcal{M}_{T,sd}^{PP}\right) \nonumber \\
+&\frac{g_1^{\pi N}}{2}\left(\mathcal{M}_{GT,sd}^{AP}+\mathcal{M}_{T,sd}^{AP}\right)
-\frac{2}{g_A^2}g_1^{NN}\mathcal{M}_{F,sd}\Bigg]\,,
\end{align}
where the QCD evolution is~\cite{Buras:2000if,Buras:2001ra,Cirigliano:2017djv},
\begin{align*}
\label{QCDrunning}
\eta(\mu_0,m_i) =\left\{
  \begin{array}{@{}ll@{}}
   \left(\frac{\alpha_s(m_i)}{\alpha_s(\mu_0)}\right)^{6/25} & m_i\leq m_{\text{bottom}} \\
\left(\frac{\alpha_s(m_{\text{bottom}})}{\alpha_s(\mu_0)}\right)^{6/25}\left(\frac{\alpha_s(m_i)}{\alpha_s(m_{\text{bottom}})}\right)^{6/23} & m_{\text{bottom}} \leq m_i\leq m_{\text{top}}\\
\left(\frac{\alpha_s(m_{\text{bottom}})}{\alpha_s(\mu_0)}\right)^{6/25}\left(\frac{\alpha_s(m_{\text{top}})}{\alpha_s(m_{\text{bottom}})}\right)^{6/23} \left(\frac{\alpha_s(m_i)}{\alpha_s(m_{\text{top}})}\right)^{2/7} & m_i \geq m_{\text{top}}
  \end{array}\right. \,,
\end{align*}
for bottom and top quark masses $m_{\text{bottom}}$ and $m_{\text{top}}$. $\alpha_s(\mu)=\frac{2\pi}{\beta_0\log(\mu/\Lambda^{(n_f)})}$, with $\beta_0 = 11-\frac{2}{3}n_f$, is the strong coupling constant, and $\alpha_s(m_Z) = 0.1179$ \cite{Workman:2022ynf}, gives $\Lambda^{(4,5,6)} \simeq \{119,\, 87,\, 43 \}$ MeV. We consider only $g^{NN}_1$ and $g_1^{\pi\pi}$, and use  $g_1^{NN}= (1+3 g_A^2)/4 $, and $g^{\pi\pi}_1=0.36$ \cite{Nicholson:2018mwc}.

The long-distance piece is computed with the interpolation formula
\begin{equation}\label{eq:fitM}
\amp^\text{(ld)}(m_i) =-\mathcal{M}(0)\frac{1}{1+m_i/m_a+(m_i/m_b)^2}\,,
\end{equation}
while short-distance part 
\begin{align}
    \amp^\text{(sd)}(m_i)=- 2 g_\nu^{NN}(m_i) \,m_\pi^2 \frac{\mathcal{M}_{F,sd}}{g_A^2}
\end{align}
contains another interpolation formula~\cite{Cirigliano:2024ccq}
\begin{equation}\label{eq:gnu_int}
g_{\nu}^{NN}(m_i) = g_{\nu}^{NN}(0) \frac{1+ a_2 \,m_i^2+ a_4\,m_i^4}{1 + |b_4|m_i^4 + |b_6|m_i^6}\,,
\end{equation}
where $g_{\nu}^{NN}(0) = -1.01\,\mathrm{fm}^2$ \cite{Cirigliano:2020dmx,Jokiniemi:2021qqv}, and the fit parameters are $a_2=8.8\text{~GeV}^{-2},\,a_4=-1.9\text{~GeV}^{-4},b_4=10\text{~GeV}^{-4}$, and $b_6=11\text{~GeV}^{-6}$. The values of $m_a$, $m_b$, and $\mathcal{M}(0)$, as well as NMEs present in the dimension-9 amplitude, for $^{136}$Xe and $^{76}$Ge, are given in \cref{tab:NME}.

\begin{table}
	\center
		\renewcommand{\arraystretch}{1.2}    
	\begin{tabular}[b]{|c|cccccccc|}    
		\hline		
		&  $\mathcal M_{F,sd}$ & $\mathcal M_{GT,sd}^{AP}$ &$\mathcal M_{GT,sd}^{PP} $&$\mathcal M_{T,sd}^{AP} $&$\mathcal M_{T,sd}^{PP}$ & $m_a$ & $m_b$ & $\mathcal{M}(0)$\\\hline
		$^{76}$Ge&-2.21 & -2.26 &0.82&-0.05&0.02 &117&218&3.4\\
		$^{136}$Xe&-1.94 &-1.99 &0.74 &0.05 &-0.02&157&221&2.7\\\hline
	\end{tabular}
	\caption{Shell-model NMEs~\cite{Menendez:2017fdf,Jokiniemi:2021qqv} and fit parameters used in \ovbb\;computations for ${}^{76}$Ge and ${}^{136}$Xe. $m_a,\,m_b$ are dimensionful, and are given in MeV here.} \label{tab:NME}
\end{table}

Several of the aforementioned NMEs, as well as ultrasoft contributions which we do not consider here, have not yet been computed for other isotopes. Nevertheless, for active neutrinos in the $m_i\rightarrow0$ limit, we require only the NME $\mathcal{M}(0)$, which is a combination of the Fermi, Gamow-Teller, and tensor components, for the long-distance part, and $\mathcal{M}_{F,sd}$ for the short-distance term. For \cref{fig:isotopes}, we pick the intermediate values $\amp^\text{(sd)} = -1.30,\,-2.00$, and $\mathcal{M}(0) = 3.3,\,3.6$ for $^{82}$Se and $^{100}$Mo respectively~\cite{Jokiniemi:2021qqv}. Note that while for all other isotopes we use the shell model determinations of parameters, the parameters we use for $^{100}$Mo are from pnQRPA, as a shell model calculation is not available.

\Cref{eq:fitM} serves as a parametric functional form the computation of the neutrino potential, which, for neutrino momentum $\vect q$ (note that the time component $q^0\ll|\vect q|$ is negligible), is given as~\cite{Dekens:2020ttz}
\begin{align}
    V_\nu(\vect q, m_i) = &-(\tau_1^+\tau_2^+)g_A^2G_F^2\left(C_\text{VLL}^{(6)}\right)^2_{ei}\left(\frac{m_i}{\vect q^2+m_i^2}\right)\nonumber\\&\times\left(-\frac{g_V^2}{g_A^2}h_F(\vect q^2) + \boldsymbol{\sigma}_1\cdot\boldsymbol{\sigma}_2\,h_{GT}(\vect q^2)+S^{(12)}h_T(\vect q^2)\right)\left[\overline{u}(p_1)P_R u^c(p_2)\right]\,,
\end{align}
where $C_\text{VLL}^{(6)}$ is the Wilson coefficient for the dimension-6 operator mediating the decay of a down quark to an up quark, an electron and an antineutrino, and the tensor operator $S^{(12)}$ is defined as $S^{(12)} = \boldsymbol{\sigma}_1\cdot\boldsymbol{\sigma}_2 - 3 (\boldsymbol{\sigma}_1\cdot\hat{\vect q} )\,(\boldsymbol{\sigma}_2\cdot\hat{\vect q})$. $h_F,\,h_{GT},\,h_T$ are respectively the Fermi, Gamow-Teller, and tensor NMEs. The integral over $\vect q$ is then performed in nuclear many-body calculations to obtain the mass-dependent long-distance amplitude.

Defining 
\begin{align}
    \Tilde{V}_\nu = V_\nu/m_i = -\frac{\kappa}{\vect q^2 + m_i^2}\,,
\end{align}
we see that 
\begin{align}
    \Tilde{V}_\nu' = &2m_i^2\frac{\kappa}{(\vect q^2 + m_i^2)^2}\,,\\
    \Rightarrow \left(1 + m_i \frac{\partial}{\partial m_i}\right)\Tilde V_\nu = &-\kappa\left(\frac{1}{\vect q^2 + m_i^2} - \frac{2m_i^2}{(\vect q^2 + m_i^2)^2}\right)\,.
\end{align}
This effectively reproduces the form of the term proportional to $\stc$ in \cref{eq:sterileamp} for $q^0\ll|\vect q|$. Given that the transition matrix element for \ovbbp\;is related to the \ovbb\;matrix element with the replacement $m_i/m_e\rightarrow \stc\Theta_{eN}^2$, and that $\kappa$ is independent of the neutrino mass, this justifies the use of the form in \cref{eq:sterileampder} for the \ovbbp\;amplitude for the long-distance contribution. 

A similar argument can be made for the short-distance contribution, where the mass-dependence enters only through the LEC $g_\nu^{NN}$, for which the functional for \cref{eq:gnu_int} is used. The same is true for the amplitude at large masses, which has an explicit $\propto1/m_i^2$ scaling, and a mass-dependence in the QCD running, which however has a very small effect.

\section{\texorpdfstring{Electron angular correlation for \ovbb}{Electron angular correlation for neutrinoless double beta decay}}
\label{app:angular}

We show here that $\mathcal{A}^\phi =\mathcal{B}^\phi$ at leading order in electron energies, i.e., neglecting the difference in their energies. Since the matrix element for \ovbbp\; reduces to that of \ovbb\;for active neutrinos, we show the equivalence for \ovbb\;instead. For simplicity, only the potential region is considered.

In Fourier representation, the amplitude is given by
\begin{align}
    \bra{e_1e_2h_f}S_\text{eff}\ket{h_i} \propto &\int d^4x\,d^4r\,\frac{d^4k}{(2\pi)^4}\frac{e^{ik\cdot r}}{k^2+i\epsilon}\bra{e_1e_2}T\left\{\overline{e}_L\left(x+r/{2}\right)\gamma^\mu\gamma^\nu e_L^c\left(x - r/2\right)\right\}\ket{0}\nonumber\\
    &\times \bra{h_f}T\left\{J_\mu(x+r/2)J_\nu(x-r/2)\right\}\ket{h_i}\,,
\end{align}
where $h_f,\,h_i$ are the final and initial nuclear states, $S_\text{eff}$ is the effective action, $J_{\mu,\nu}$ are the quark currents, $k$ is the virtual neutrino momentum, and $T$ represents time-ordering. Neglecting the difference in electron momenta, this becomes~\cite{Cirigliano:2017tvr}
\begin{align}
    \bra{e_1e_2h_f}S_\text{eff}\ket{h_i} \propto &\int d^4x\,d^4r\,\frac{d^4k}{(2\pi)^4}\frac{e^{ik\cdot r}}{k^2+i\epsilon}\bra{e_1e_2}\overline{e}_L\left(x\right) e_L^c\left(x \right)\ket{0}g^{\mu\nu}\nonumber\\
    &\times \bra{h_f}T\left\{J_\mu(x+r/2)J_\nu(x-r/2)\right\}\ket{h_i}\,.
\end{align}
The operators can be made independent of the arguments using translation invariance -- $\mathcal{O}(x) = e^{ix\cdot P}\mathcal{O}(0)e^{-ix\cdot P}$. The integral over $x$ then gives the delta function for momentum conservation. Inserting the a complete set of states, $\sum_n\ket{n(\vect q)}\bra{n(\vect q)}$, between the quark currents, the $\vect r$ integrals give $\Theta(r^0)\delta^{(3)}\left(\vect q -\vect k - \frac{\vect p_i + \vect p_f}{2}\right) + \Theta(-r^0)\delta^{(3)}\left(\vect q +\vect k - \frac{\vect p_i + \vect p_f}{2}\right)$ with the Heaviside function $\Theta(a)$, while the $r^0$ and $k^0$ integrals give the energy denominator $E_n - \frac{E_i + E_f}{2} + |\vect k|$. The hadronic part is thus independent of electron momenta at this order, and the electron angular correlation depends only on the leptonic part of the matrix element.

The lepton line, without the neutrino propagator, is $L^{\mu\nu} = \overline{u}(p_{e_1})\,P_R\,u^c(p_{e_2})g^{\mu\nu}$. For the squared matrix element, we have
$\frac12\sum_\text{spins}L^{\mu\nu}L^{\rho\sigma}g_{\mu\nu}g_{\rho\sigma} = 4\,p_{e_1}\cdot p_{e_2} = 4(E_{e_1} E_{e_2} - \vect p_{e_1}\cdot \vect p_{e_2})$. To account for the Coulomb interaction of the electrons in the nuclear field, the electron spinors can be modified using the factors $g_{-1}$ and $f_1$ as~\cite{Deppisch:2020mxv}
\begin{equation}
    u_e^{(s)} = \left(\begin{matrix}
        g_{-1}(E_e)\,\chi^{(s)}\\ f_{1}(E_e)\left(\boldsymbol\sigma\cdot \hat{\vect p}_e\right) \chi^{(s)}
    \end{matrix}\right)\,,
\end{equation}
where the superscript denotes the spin, and the three-momentum $\hat{\vect p}_e$ is normalised to its magnitude. This results in an effective replacement
\begin{align}
    E_e\rightarrow &\,E_e \left(g_{-1}^2(E_e) + f_1^2(E_e)\right)\\
    \vect p_e \rightarrow &\, 2E_e\,f_1(E_e)\,g_{-1}(E_e) \,\hat{\vect p}_e\,,
\end{align}
which leads to 
\begin{align}
    \frac12\sum_\text{spins}L^{\mu\nu}L^{\rho\sigma}g_{\mu\nu}g_{\rho\sigma} \propto &\, E_{e_1}E_{e_2}\Bigg[\left(g_{-1}^2(E_{e_1}) + f_1^2(E_{e_1})\right)\left(g_{-1}^2(E_{e_2}) + f_1^2(E_{e_2})\right)\nonumber\\
    & - \hat{\vect p}_{e_1}\cdot \hat{\vect p}_{e_2}\left(2f_1(E_{e_1})\,g_{-1}(E_{e_1})\right)\left(2f_1(E_{e_2})\,g_{-1}(E_{e_2})\right)\Bigg]\,,
    \label{eq:fermi_spinsum}
\end{align}
where $\hat{\vect p}_{e_1}\cdot \hat{\vect p}_{e_2} = \cos\theta$ gives the angle between the electrons. The factors can now be identified as the Fermi function $F(Z,E_e)$ and the function $E(Z,E_e)$ shown in \cref{eq:fermi_mod}.

From the definition of the differential rate in \cref{eq:angcorrrate}, we can read off $\Lambda^\phi$ and $\Gamma^\phi$ with the $\theta$-independent and -dependent terms in \cref{eq:fermi_spinsum}. Hence, we see that, at this order, $\mathcal{A}^\phi=\mathcal{B}^\phi$, with the only difference in the prefactors for $\Lambda^\phi$ and $\Gamma^\phi$ being the functions $F(Z,E_e)$ and $E(Z,E_e)$.

\bibliographystyle{utphys}
\bibliography{references.bib}

\end{document}